\documentclass[reprint, prd, superscriptaddress, longbibliography, tightenlines, nofootinbib, eqsecnum, floatfix, notitlepage,twocolumn]{revtex4-2}
\pdfoutput=1

\usepackage[utf8]{inputenc}
\usepackage{euscript}
\usepackage{epsfig}
\usepackage{graphics}
\usepackage{graphicx}
\usepackage{amsmath}
\usepackage{amssymb}
\usepackage{amsfonts}
\usepackage{bm}
\usepackage[usenames,dvipsnames,svgnames,table]{xcolor}
\usepackage{xspace}
\usepackage{times}
\usepackage{appendix}
\usepackage{lipsum}
\usepackage[nolist,nohyperlinks]{acronym}
\usepackage{float}
\usepackage{simplewick}
\usepackage{tabularx}
\usepackage{booktabs}
\usepackage{accents}

\usepackage{natbib, ifthen}
\usepackage{hyperref}
\usepackage{comment}

\renewcommand{\vr}{\boldsymbol{r}}
\newcommand{\vl}{\ensuremath{\boldsymbol{l}}\xspace}

\newcommand{\vL}[0]{\ensuremath{\boldsymbol{L}}\xspace}

\newcommand{\jcap}{J. Cosmol. Astropart. Phys.}

\newcommand{\isdraft}[1]{}

\newcommand{\JC}[1]{{\isdraft{\color{purple} JC: #1}}}
\newcommand{\AL}[1]{{\isdraft{\color{blue} AL: #1}}}

\newcommand{\hn}[0]{{\hat n}}

\newcommand{\Ml}[0]{\ensuremath{\mathsf{M}}} 
\newcommand{\av}[1]{\left \langle #1 \right \rangle}
\newcommand{\avM}[1]{\left \langle #1 \right \rangle_{\Ml}}
\newcommand{\evenweight}{[\Delta^L_{0\Ml}]^2}
\newcommand{\dlp}[3]{\ensuremath{\Delta^{#1}_{#2#3}}}
\newcommand{\T}[0]{\mathcal T}
\newcommand{\U}[0]{\mathcal U}

\graphicspath{{Figures/}}

\newcommand{\begm}{\begin{pmatrix}}
\newcommand{\enm}{\end{pmatrix}}
\newcommand{\threej}[6]{{\begm #1 & #2 & #3 \\ #4 & #5 & #6 \enm}}

\newcommand\ba{\begin{eqnarray}}
\newcommand\ea{\end{eqnarray}}
\newcommand\bea{\begin{eqnarray}}
\newcommand\eea{\end{eqnarray}}

\newcommand\be{\begin{equation}}
\newcommand\ee{\end{equation}}
\newcommand{\la}{\langle}
\newcommand{\ra}{\rangle}
\renewcommand{\d}{\text{d}}

\newcommand{\lplus}{\ell}
                               
\newcommand{\clmat}[0]{\textbf{C}}

\newcommand{\Cgrad}{\tilde{C}}

\begin{document}

\title{Spherical bispectrum expansion and quadratic estimators}

\author{Julien Carron}
\affiliation{Universit\'e de Gen\`eve, D\'epartement de Physique Th\'eorique et CAP, 24 Quai Ansermet, CH-1211 Gen\`eve 4, Switzerland}

\author{Antony Lewis}
\affiliation{Department of Physics \& Astronomy, University of Sussex, Brighton BN1 9QH, UK}

\date{\today}

\begin{abstract}
We describe a general expansion of spherical (full-sky) bispectra into a set of orthogonal modes. For squeezed shapes, the basis separates physically-distinct signals and is dominated by the lowest moments. In terms of reduced bispectra, we identify a set of discrete polynomials that are pairwise orthogonal with respect to the relevant Wigner 3j symbol, and reduce to Chebyshev polynomials in the flat-sky (high-momentum) limit for both parity-even and parity-odd cases. For squeezed shapes, the flat-sky limit is equivalent to previous moment expansions used for CMB bispectra and quadratic estimators, but in general reduces to a distinct expansion in the angular dependence of triangles at fixed total side length (momentum).
We use the full-sky expansion to construct a tower of orthogonal CMB lensing quadratic estimators and construct estimators that are immune to foregrounds like point sources or noise inhomogeneities. In parity-even combinations (such as the lensing gradient mode from $TT$, or the lensing curl mode from $EB$) the leading two modes can be identified with information from the magnification and shear respectively, whereas the parity-odd combinations are shear-only. Although not directly separable, we show that these estimators can nonetheless be evaluated numerically sufficiently easily.
\end{abstract}

\maketitle
 \tableofcontents

\section{Introduction}
Cosmological observations are made on the celestial sphere, and appear to be broadly consistent with statistical isotropy. The CMB and large-scale structure appear to be nearly Gaussian on large scales, suggesting that the primordial fluctuations were Gaussian or only perturbatively non-Gaussian. However, since the information in a Gaussian field is limited to the power spectrum, a lot of the available information potentially lies in the non-Gaussian signal. At perturbative order, this information lies mainly in the bispectrum (three-mode correlation) or trispectrum (connected four-mode correlation). The dominant blackbody trispectrum in the CMB is due to lensing, and can equivalently be thought of as the power spectrum of a quadratic estimator for the lensing defection field responsible for the signal. The correlation between the quadratic estimator and the statistically isotropic lensing field responsible for the trixspectrum has the form of a bispectrum, and this bispectrum contains the key physical signal (but is not directly observable). This paper aims to present general decomposition of bispectra observed on the celestial sphere, which can be used to isolate and test distinct physical contributions to both primordial and late-time non-Gaussianity.

There is already a well-developed technology for decomposing bispectra into sums of terms, often with the objective of making the individual terms fast to evaluate by means of separable estimators~\cite{Komatsu:2003iq,Fergusson:2008ra}, or to split the signal into parts that can be tested for self-consistency~\cite{Munshi:2009ik}. Decompositions can also be constructed to easily remove terms that are likely to be contaminated (e.g. by foregrounds), and/or where the signal is contained in only a small number of terms~\cite{Schaan:2018tup,Qu:2022qie}.
Here, we are more interested in the latter approach, showing that even non-separable estimators can be evaluated quite easily using modern techniques. Specifically, we aim to generalize the flat-sky bispectrum angular shape decomposition of Refs.~\cite{Lewis:2011au,Pearson:2012ba,Schaan:2018tup} to observations on the sphere, developing the approach initially suggested in the appendix of Ref.~\cite{Pearson:2012ba}. Unlike Ref.~\cite{Qu:2022qie}, which successfully implements the approximate flat-sky decomposition covariantly on the sphere, we aim for a fully curved-sky approach that is in principle exact and can be used to describe the lowest multipoles self-consistently.
The multipole shape decomposition helps to isolate physically distinct effects, for example sub-horizon dynamical tidal effects from angle-independent super-horizon scalar local non-Gaussianity, or lensing shear from convergence and point sources. It also gives a basis in which a few lowest multipoles carry most of the signal when the signal is predominantly in squeezed shapes (with two wavenumbers much larger than the other). The presence of a quadrupolar (or higher) shape dependence on super-horizon scales of primordial origin would be a signal of higher spin fields during inflation~\cite{Shiraishi:2013vja,Arkani-Hamed:2015bza}.

   \begin{figure}
    \centering
    \includegraphics[width=\columnwidth]{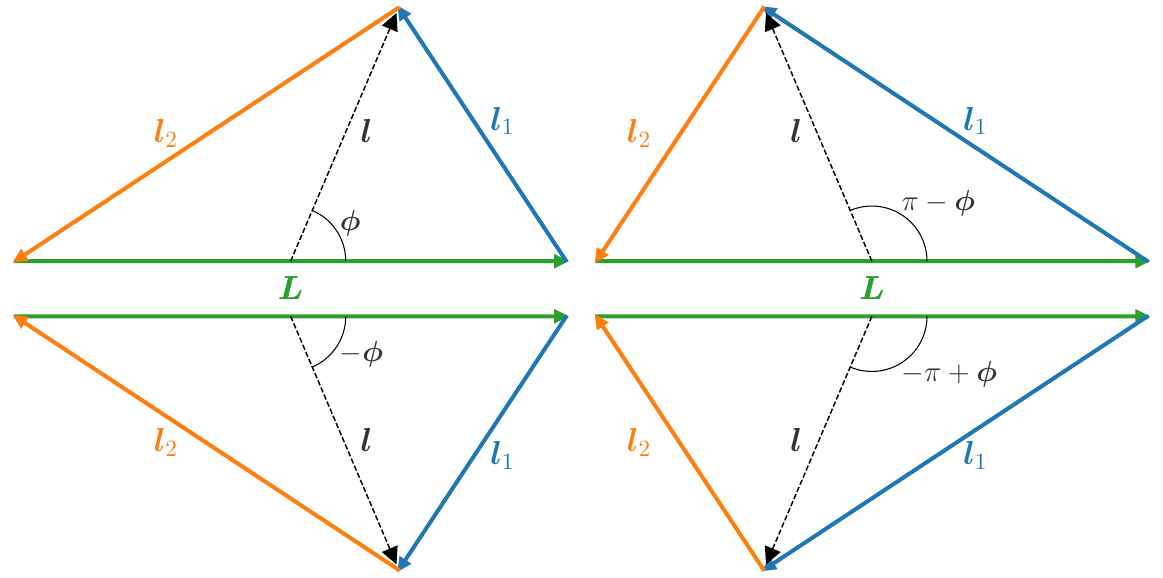}
    \caption{In the flat sky the three wavevectors $\vL, \vl_1, \vl_2$ of a bispectrum form a triangle. The triangle can also be described by $L$, $l\equiv |\vl|$ and $\phi$ where $\vl\equiv  \frac 12 (\vl_1-\vl_2)$. The wavevectors are colored to represent the three possibly-different fields making up the bispectrum. The upper and lower triangles are related by a reflection, and hence are equivalent in the parity-even case, so the bispectrum must be even in $\phi$ (and hence no $\sin(m\phi)$ dependence in  the bispectrum shape). If one or three of the fields are parity odd, the lower triangles have opposite sign bispectrum, so the dependence in $\phi$ is then odd. Triangles on the left and right are related by $\ell_1 \leftrightarrow \ell_2$. If those two fields are the same, the bispectrum is symmetric in $\ell_1, \ell_2$ so the shape dependence must be symmetric about $\phi=\pi/2$; i.e., the mode expansion only has even $m$ modes, and has even dependence on $\Ml\equiv \ell_2-\ell_1$. A component of a mixed-field bispectrum that is antisymmetric in $\ell_1, \ell_2$ would give odd $m$ modes.
    In the common simple case of a bispectrum of only one parity-even field, all four triangles are equivalent and only moments of $\cos(m\phi)$ with $m$ even are allowed.
    \label{fig:triangles}
    }
\end{figure}

In the flat sky, a 2D statistically-homogeneous bispectrum can be described by three wavevectors forming a triangle, where for a statistically isotropic signal the bispectrum only depends on the lengths of the sides $L, \ell_1\equiv |\vl_1|, \ell_2\equiv |\vl_2|$ (and possibly a sign in the case of a parity odd bispectrum with mixed fields). Alternatively, we can describe the shape via the length of one side, $L$, $l \equiv \frac 12 |\vl_1-\vl_2|$ and the angle $\phi$ between $\vL$ and $\vl$; see Fig.~\ref{fig:triangles}. The dependence of the bispectrum on the angle $\phi$ can be expanded into multipole moments $e^{im\phi}$, where $m$ is the multipole for the shape-dependence of the bispectrum. For parity even bispectra of a single field, $m\ge 0$ is even by symmetry of how the angle is defined. In squeezed shapes, the bispectrum is dominated by the lowest multipoles, describing monopole ($m=0$) or quadrupole ($m=2$) dependence from distinct physical effects (see Ref.~\cite{Lewis:2011au} for a review and diagrams). Alternatively, we could describe the triangle by $L$, $(\ell_1+\ell_2)/2$ and $\ell_2-\ell_1$,  giving a different flat-sky decomposition sharing the same key properties. We shall show that this generalizes nicely to the full sky, and in the flat-sky limit gives a new distinct multipole expansion to describe shape dependence at fixed total side length. 

On the curved sky a statistically isotropic and parity invariant bispectrum is defined as
\begin{equation}\label{eq:bispec}
B^{ijk}_{L \ell_1 \ell_2} \equiv \sum_{M m_1 m_2} \threej{L}{\ell_1}{\ell_2}{M}{m_1}{m_2} \la a^i_{L M}  a^j_{\ell_1 m_1} a^k_{\ell_2 m_2}\ra ,
\end{equation}
where $a^i_{L M}$ is the spherical-harmonic multipole of field $i$.
We consider quadratic estimators and bispectra together in the language of bispectra for consistency. A quadratic estimator response kernel is also of the form of a bispectrum, since for an underlying Gaussian field $\psi_{LM}$ that is being reconstructed the kernel is of the form~\cite{Lewis:2011fk}
\begin{eqnarray}
f^{\psi a b}_{L \ell_1 \ell_2} &=& \sum_{Mm_1 m_2} \threej{L}{\ell_1}{\ell_2}{M}{m_1}{m_2}  \left\la \frac{\delta}{\delta \psi_{LM}^*}\left( a_{\ell_1 m_1} b_{\ell_2 m_2}\right) \right\ra \nonumber
\\&=& [C^{\psi\psi}_{L}]^{-1} B^{\psi a b}_{L \ell_1 \ell_2}.
\label{lenskernel}
\end{eqnarray}
The bispectrum multipoles $L, \ell_1, \ell_2$ are restricted by the usual triangle constraints $|\ell_1-\ell_2| \le L \le |\ell_1+\ell_2|$.
This paper is based on the simple observation that sums over the bispectrum take a particularly convenient form when expressed in the variables
\begin{equation}
	\ell \equiv  \frac 12(\ell_1 + \ell_2), \quad \Ml \equiv \ell_2 - \ell_1.
\end{equation}
Using the new integers $\Ml$ and $2\ell$, sums over the multipoles can be written 
\begin{equation}
	\sum_{L \ell_1 \ell_2} = \sum_L \sum_{2\ell \ge L} \sum_{\Ml = -L}^L,
\end{equation}
where the limits on the right-hand side follows from the triangle conditions on the multipoles. 
One motivation for this choice of variable is Regge symmetry of the Wigner 3j symbol~\cite{Regge:1958aa},
\begin{equation}\label{eq:Regge}
	\begin{pmatrix}
		L & \ell_1 & \ell_2 \\ 0 & 0 & 0
	\end{pmatrix} = \begin{pmatrix}
		L & \ell & \ell  \\ -\Ml & \Ml/2 & \Ml/2 
	\end{pmatrix}.
\end{equation}
Here $\Ml$ naturally appears as a quantum number associated to $L$, obtained from the addition of two now identical angular momenta (see also Eq.~\eqref{eq:Regge2}).
The sum over $L, \Ml$ now also has the same form as the sum over spherical harmonic multipole coefficients $a_{L \Ml}$ for $2\ell \ge L$.
This suggests defining a set of bispectrum ``configuration space'' multipoles for each $\ell$ (dropping field indices for simplicity for now, see Appendix~\ref{ap:multifield} for general form)
\be \label{eq:BLM}
B_{\lplus,L\Ml} \equiv (-1)^{\lfloor{(L-\Ml)/2}\rfloor} B_{L (\lplus-\Ml/2) (\lplus+\Ml/2)}.
\ee
Note that the real part of $B_{\lplus,L\Ml}$ is parity even in the fields, the imaginary part is parity odd. The ($-1$) factor ensures that $B_{\lplus,L\Ml}^* = (-1)^\Ml B_{\lplus, L-\Ml}$ so that they have the symmetries of the spherical harmonic multipoles. For any fixed $\lplus$ or $L$, this defines a spherical map that can be used to visualize the bispectrum shape dependence.

In the flat sky case, the angular dependence of the bispectrum shape can be expanded in $e^{im\phi}$, which are the eigenfunctions of rotations in the plane of the 
bispectrum. In the curved sky, we can instead expand the bispectrum into a complete set of discrete eigenmodes.

Eigenmodes of rotations about the $x$-axis can be obtained from the eigenvalue equation\footnote{We use eigenmodes of rotation about $x$ rather than $y$, because only the $x$ eigenmodes have symmetry consistent with $B_{\lplus,L\Ml}^* = (-1)^M B_{\lplus, L-\Ml}$. Instead expanding $B_{L (\lplus-\Ml/2) (\lplus+\Ml/2)}$ directly in eigenmodes of $y$ rotations would give the same overall result up to a phase.}
\ba
\label{xeigs}
\sum_{M'} X^L_{MM'}(\theta) d^L_{mM'}(\pi/2)= e^{-im\theta} d^L_{mM}(\pi/2),
\ea
which follows from the well-known result for calculating a rotation about $x$ as a combination of a rotation about $z$ with two rotations by $\pi/2$~\cite{Risbo:1996aa}:
\begin{multline}
D^L_{MM'}(\pi/2, \theta, -\pi/2) \equiv X^L_{MM'}(\theta)
\\
= \sum_{m=-L}^{L} d^L_{mM}(\pi/2) e^{-i m\theta} d^L_{mM'}(\pi/2).
\end{multline}

   \begin{figure}
    \centering
    \includegraphics[width=\columnwidth]{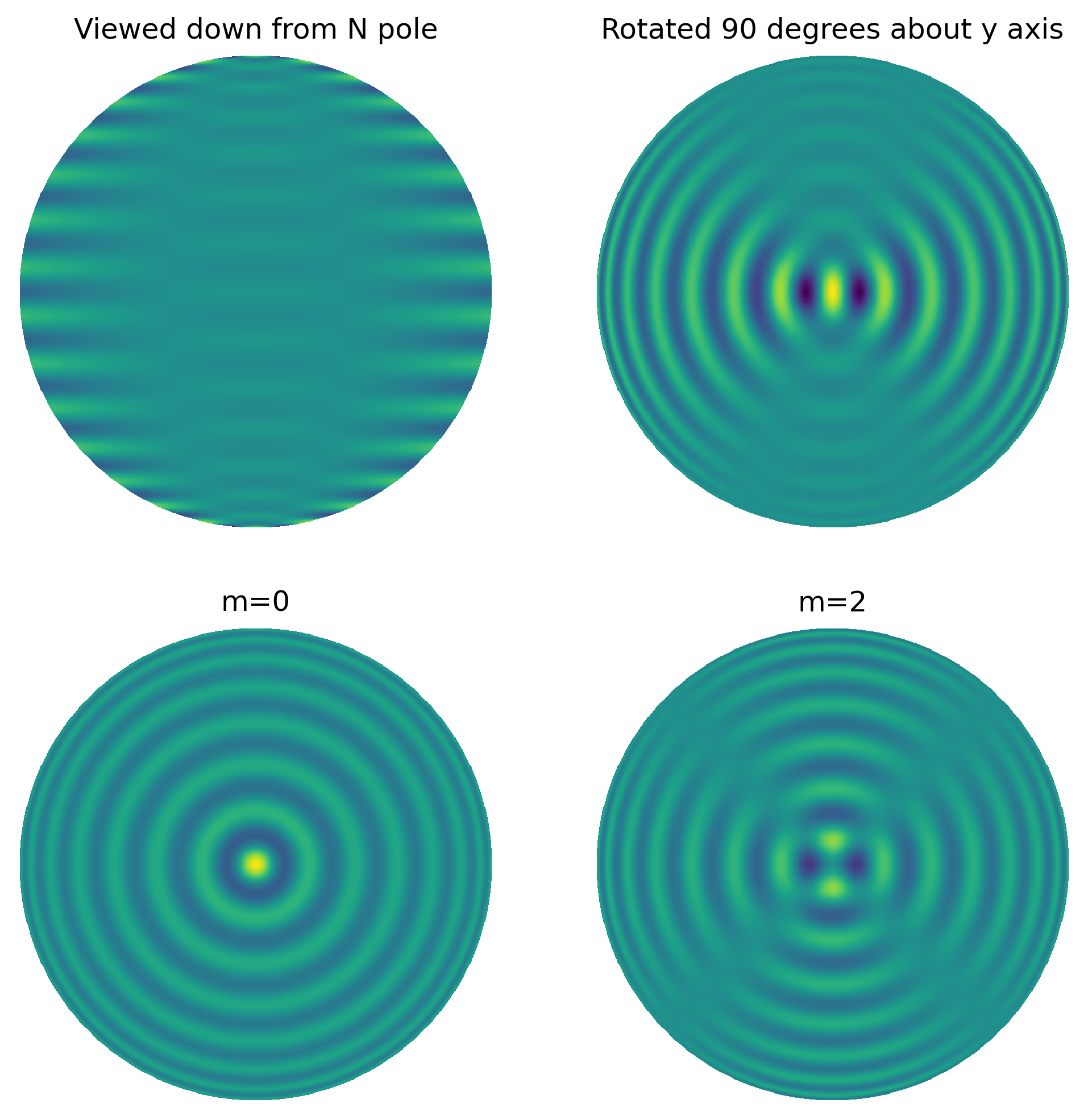}
    \caption{A moment expansion example for the weighted CMB lensing temperature response function. The top left figure shows the configuration space $\ell,\Ml$ map for fixed $L=30$ (Eq.~\eqref{eq:BLM}, summed over $50 \le \ell\le 1500$). The top right figure shows this map rotated by $\pi/2$ about the $y$ axis, which rotates the coordinates such that the spherical $m$-multipoles are now the moments of interest (Eq.~\eqref{eq:Bmintro}). The lower two maps show the dominant $m=0$ and $m=2$ components, corresponding to lensing magnification and shear. The response functions are $\sqrt{C_\ell}$-weighted (see Sec.~\eqref{sec:QE}) with no noise.
    \label{fig:rots}
    }
\end{figure}

The eigenmodes of rotations about $x$ are simply rotations about $y$ by $\pi/2$, which we denote $\Delta^L_{mM} \equiv  d^L_{mM}(\pi/2)$. Expanding $B_{\lplus,L\Ml}$ into $m$ eigenmodes is therefore just equivalent to rotating the bispectrum configuration space multipoles by $\pi/2$:
\be \label{eq:Bmintro}
B_{\lplus L}^{(m)} \equiv \sum_{\Ml =-L}^L \Delta^L_{m\Ml} B_{\lplus, L\Ml}. 
\ee
A general bispectrum can therefore be decomposed uniquely into a sum over different $m$ moments. It is less obvious why this is useful, but we shall show that for squeezed bispectra these moments are a natural spherical generalization of the flat sky angular multipoles, and are well described by a small number of low even values of $m$ (see lensing illustration in Fig.~\ref{fig:rots}). In particular, for squeezed shapes with $L \ll \ell_1,\ell_2$, the $\Ml$-dependence of the Gaunt integral that enters standard scalar bispectra becomes $\propto \Delta^{L}_{0\Ml}$, in which case $B_{\lplus L}^{(m)}\propto \delta_{m0}$ if the reduced bispectrum has no strong shape dependence.

The original bispectrum can be recovered from the moments using
\begin{multline} \label{eq:Binv}
B_{L (\ell-\Ml/2)(\ell+\Ml/2)} =  (-1)^{\lfloor{(L-\Ml)/2}\rfloor}\sum_{m}\Delta^L_{m\Ml} B^{(m)}_{\ell L} \\
=(-1)^{\lfloor{(L-\Ml)/2}\rfloor}\sum_{m\ge 0} (2 - \delta_{m0}) \Delta^L_{m\Ml} B^{(m)}_{\ell L},
\end{multline}
and we consider $m$ to be positive or zero from now on. For this simple case with one field, $m$ must be even; in general, mixed field configurations that are not symmetric under interchange of $\ell_1$ and $\ell_2$ (and hence $\Ml \leftrightarrow -\Ml$) can also give rise to odd $m$.

In the following sections we show that, for parity even bispectra, this expansion is equivalent to an expansion in a set of orthogonal polynomials which are a quantized version of the Chebyshev polynomials (and reduce to them in the flat-sky limit). For squeezed shapes, this is equivalent to expanding the reduced bispectrum in a set of orthogonal polynomials.
We also give the equivalent expansion for parity odd bispectra, and discuss the application to CMB lensing quadratic estimators, and the choice of re-weighting functions that can be used before doing the multipole decomposition depending on whether the objective is to obtain orthogonal estimators or to project monopolar contaminating signals.
We then take the flat-sky (high-momentum) limit, giving a new multipole decomposition for flat sky bispectra, and discuss equivalence in the squeezed limit.

\section{Polynomial bispectrum expansion}
We now show the equivalence of this configuration space bispectrum rotation to an expansion of a reduced bispectrum (as made precise later on) in a set of Chebyshev-like polynomials, which are orthogonal with respect to the relevant Wigner-3j symbols. 
\subsection{Parity-even}
When the bispectrum is a correlator of three fields that are all parity invariant, or contains an even number of parity odd fields, we say the bispectrum is parity even. A parity-even bispectrum must have $L+\ell_1+\ell_2$ even for overall statistical parity invariance; likewise, for parity-odd, the sum must be odd. We first focus on the parity-even case, where the reduced bispectrum\footnote{In quantum mechanics language, $B^{ijk}_{L \ell_1 \ell_2}$ already corresponds to the reduced matrix element since it is defined using the symmetries of the Wigner-Eckart theorem.} $b_{L\ell_1 \ell_2}$ is usually defined as~\cite{Komatsu:2001rj}
\begin{equation}\label{eq:reducedB}
	b_{L \ell_1 \ell_2} 	\begin{pmatrix}
		L & \ell_1 & \ell_2 \\ 0 & 0 & 0
	\end{pmatrix}\sqrt{\frac{(2L+ 1) (2\ell_1 + 1)(2\ell_2 + 1)}{4\pi}} \equiv B_{L \ell_1 \ell_2} 
\end{equation}
The Wigner 3j symbol is never zero for $L+\ell_1+\ell_2$ even and multipoles satisfying the triangle condition, so we can always make this definition. The form of Eq.~\eqref{eq:reducedB} is well motivated, as this combination of the 3j symbol appears in the Gaunt integral of three spherical harmonics, and $b_{L \ell_1 \ell_2}$ agrees in the appropriate limit with the flat-sky result~\cite{Hu:2000ee}.
More generally, when the underlying fields have non-zero spin, different factors can appear, but they can always be re-written by factoring out 3j dependence of Eq~\ref{eq:reducedB} as we discuss further in Appendix~\ref{app:QEs}.

The Wigner 3j element can be written explicitly in the following form\footnote{The sign before squaring is $(-1)^s$.}~\cite{Adams3Pn}
\begin{align}\label{eq:3j2} 
			&\begin{pmatrix}
		L & \ell_1 & \ell_2 \\ 0 & 0 & 0
	\end{pmatrix}^2 = \frac 1 {2s + 1} \frac{A(s- L)A(s- \ell_1)A(s - \ell_2)}{A(s)} \\
	&=\frac{1}{2\ell + 1 + L} \frac{A((2\ell -L)/2)}{A((2\ell + L)/2)}A( (L + \Ml)/2)A( (L - \Ml)/2) \nonumber
\end{align}
with $s \equiv \frac 12 \left( L  + \ell_1 + \ell_2 \right)$, and
\begin{align}
	A(n)& \equiv  \frac 1 {4^n} \begin{pmatrix}
		2n \\ n 	\end{pmatrix} \sim \frac{1}{\sqrt{\pi n}}\left( 1 - \frac 1{8n} + \cdots \right). \label{eq:asymp}
\end{align}
The bottom line in Eq.~\eqref{eq:3j2} shows that the 3j symbol is separable into a function of $2\ell$ and $L$, and another function of $L$ and $\Ml$ only~\cite{Pearson:2012ba}. In Eq.~\eqref{eq:asymp}, we also made explicit the high $n$ behaviour (`flat-sky limit') for future convenience. 

The separable $L,\Ml$-dependent factor can be written in terms of the Wigner small d-matrix  (for $L+\Ml$ even,~\citet[][p113]{AngularMom})
\begin{multline}
\Delta^L_{0\Ml}\equiv d^{L}_{0\Ml}(\pi/2) \\
= (-1)^{(L-\Ml)/2} \sqrt{A((L + \Ml)/2)A( (L - \Ml)/2)}.
\end{multline}

When constructing Fisher estimates in a null hypothesis, or quantifying overlap between bispectra, squared bispectra appear in the sum over multipoles.  The separability implies that it is possible to build discrete polynomials of degree $m$ in $\Ml$, $\mathcal T_m^L(\Ml)$, that are independent of $2\ell$ and pairwise orthogonal with respect to the squared 3j symbol:
\begin{equation}
\begin{split}
	&\sum_{\Ml = -L}^L \begin{pmatrix}
		L & \ell_1 & \ell_2 \\ 0 & 0 & 0
	\end{pmatrix}^2 \mathcal T_m^L(\Ml)  \mathcal T_{m'}^L(\Ml) \\\propto &\sum_{\Ml =-L}^L  \evenweight \mathcal T_m^L(\Ml)  \mathcal T_{m'}^L(\Ml) \propto \delta_{mm'}.
\end{split}
\end{equation}
We sometimes write for $\Ml$-averages
\begin{equation}
	\avM{\cdots} = \sum_{\Ml=-L}^L \evenweight  (\cdots)
\end{equation}
when the context is clear.

Introducing the variables  $x= \Ml / L \equiv \cos \varphi$, it follows from \eqref{eq:asymp} that in the flat-sky limit ($L,\Ml$ large) the weight function becomes\footnote{In general we can define  $\cos\varphi \equiv 2\Ml/(2L+1)$. Although $|\Ml|\sim L$ can give arguments of $A$ near zero, these become a negligible fraction of the modes for large $\ell, L$.}
\begin{equation}
	 \evenweight \rightarrow \frac 1 \pi \frac 2 {L} \frac1{\sqrt{1 - x^2}}.
\end{equation}
The factor $2/L$ is just the spacing $dx$ between two $x$ points (recall that $\Ml$ always jumps by two units for fixed parity).
Hence, treating $x$ as a continuous variable in the flat-sky limit, we always recover independently of $L$ the same weight $ \propto dx /\sqrt{1 - x^2}$, or, equivalently, uniform weighting $d\varphi$ in $\varphi$. The orthogonal polynomials in $x$ associated to this weight function are the Chebyshev polynomials of the first kind, $T_m(x) \equiv \cos (m \varphi)$. It follows that we can fix the normalization of $\mathcal T^L_m(M)$ such that they all tend to $T_m(\Ml/L)$ in the flat-sky limit.

In Appendix~\ref{app:poly} we give very concise expressions for these polynomials and further discuss some of their properties. In particular, we show that they are given by
\begin{equation}
	\mathcal T^L_m(\Ml) = \frac{\Delta^L_{m \Ml}}{\Delta^L_{0\Ml}}. 
\end{equation}
In order to build the bispectrum expansion, we can: 
\begin{enumerate}
	\item Remove from $B$ the $\Ml$-dependency of the Wigner symbol. This is equivalent to dividing by $|\Delta^L_{0\Ml}|$.
	\item Project $B / |\Delta^L_{0\Ml}|$ onto the polynomial basis.
\end{enumerate}
For a parity-even bispectrum these operations result in
\begin{equation} \label{eq:Bexp}
	B^{(m)}_{L\ell} = \avM{\frac{B_{L \ell_1 \ell_2}}{|\Delta^L_{0\Ml}|} \mathcal T^L_m(\Ml)} =\sum_{\Ml = -L}^L \Delta^L_{m \Ml} B_{\ell, L \Ml}
\end{equation}
with $B_{\ell, L\Ml}$ as defined in Eq.~\eqref{eq:BLM}. This is, as already advertised in the introduction, a rotation by $\pi/2$ of the spherical map with spherical harmonic coefficients $B_{\ell, L\Ml}$ around the $y$-axis.

The inclusion of the prefactor $1/|\Delta^L_{0\Ml}|$, which is necessary for orthogonality of the expansion with respect to the $\Ml$-average, is equivalent to expanding the reduced bispectrum in $\mathcal T^L_m(\Ml)$ up to the factors $\sqrt{(2\ell_1 + 1)(2\ell_2 + 1)}$ in Eq.~\eqref{eq:reducedB} (aside from others factors that are irrelevant since they depend only on $\ell$ and $L$). 
For squeezed shapes (low $L$) and reasonable $h$, functions of the form $h_{\ell_1}h_{\ell_2}$ do not affect the expansion up to $(L /\ell)^2$ corrections. More generally, one is free to expand a rescaled bispectrum as most fit for purpose.

For example, consider the simplest case of a bispectrum induced by unclustered point sources. This has a perfectly constant reduced bispectrum 
\begin{equation}
b^{\rm PS}_{L \ell_1 \ell_2} \equiv b^{\rm PS},
\end{equation}
so that we might want to define this to be a perfect monopole. This is not the case for non-squeezed triangles in Eq.~\eqref{eq:Bexp}, precisely because of the $h_{\ell_1}h_{\ell_2}$ factors. However, if we expand the rescaled version
\begin{equation}
	\frac{B_{L\ell_1 \ell_2}}{h_{\ell_1}h_{\ell_2}}, \quad h_\ell
	\equiv \sqrt{2\ell + 1},
\label{ps_scale}
\end{equation}
instead of $B_{L\ell_1 \ell_2}$, then the resulting expansion is always a pure monopole. 

For another example of well-motivated rescaling choice, consider the signal to noise (or Fisher information) at $L$ for these bispectra. In the simple case of identical fields in the null hypothesis of Gaussian fields, the Fisher information for $B$ as defined by Eq.~\eqref{eq:Bmintro} is
\begin{equation} \label{eq:FLB}
	F_L(B) = \frac 1 6 \sum_{\ell_1 \ell_2} \frac{B^2_{L\ell_1\ell_2}}{C^{\rm tot}_L C^{\rm tot}_{\ell_1} C^{\rm tot}_{\ell_2}}.
\end{equation}
Hence, now expanding 
\begin{equation}
\label{Bscale}
	\frac{B_{L\ell_1 \ell_2}}{h_{\ell_1}h_{\ell_2}}, \quad h_\ell
	\equiv \sqrt{C_\ell^{\rm tot}}
\end{equation}
will cause the Fisher information to collapse to a sum of squares at any $L, \ell$, not just in the squeezed limit. Later when dealing with quadratic estimators we will include both rescalings.

\subsection{Parity-odd}
Let us now discuss parity-odd bispectra,  where there is no very standard definition of a reduced bispectrum.
However, as demonstrated in the appendix, we find that our results can be extended naturally to this case.

We define our quantized polynomials in the odd case as
\begin{equation}
	\mathcal U^L_m(\Ml) \equiv \frac{\Delta^L_{m\Ml}}{\Delta^L_{1\Ml}},\quad m = 1, \cdots, L.
\end{equation}
They are orthogonal with respect to the weights
\begin{equation}
	\left[\Delta^L_{1\Ml}\right]^2 = \frac{L^2-\Ml^2}{L(L+1)}\left[ \Delta^{L-1}_{0\Ml} \right]^2
\end{equation}
summed over $L+\Ml$ odd.
In the flat-sky limit these polynomials tend to the Chebyshev polynomials of the second kind,
\begin{equation}
	\mathcal U^L_m(\Ml) \rightarrow U_{m-1}(x) = \frac{\sin m \varphi}{\sin \varphi},
\end{equation}
as desired for odd-parity bispectra. The parity-odd weights take the form of the parity-even weights (but at $L-1$) with an additional angular factor of $L^2-\Ml^2= L^2\sin^2\varphi$, which makes the total weight reduce in the flat-sky limit to that corresponding to $U(x)$ (i.e. $\propto\sin \varphi$). These weights appear naturally in many bispectra. For example, in appendix~\ref{app:QEs} we discuss the lensing quadratic estimators and show that all the odd-parity lensing estimators include the $\Ml$-dependent geometric factor
\begin{equation}\label{eq:oddreduc}
	\begin{pmatrix}
		L-1 & \ell_1 & \ell_2 \\ 0 & 0 & 0 
	\end{pmatrix} \sqrt{L^2 - \Ml^2} \propto  (-1)^{\lfloor{(L-\Ml)/2}\rfloor}\Delta_{1\Ml}^L.
\end{equation}
As we discuss there, this choice of reduction is well-motivated for any odd bispectra for which the $L$-mode can be treated as Gaussian. With this in place, the expansion can proceed in the same manner as the parity-even case:
\begin{enumerate}
	\item Extracting a factor $|\Delta^L_{1\Ml}|$
	\item Projecting onto the $\mathcal U^L_m(\Ml)$ polynomial basis.
\end{enumerate}
We get
\begin{equation}\label{eq:Bexpodd}
		B^{(m)}_{L\ell} = \avM{\frac{B_{L \ell_1 \ell_2}}{|\Delta^L_{1\Ml}|} \mathcal U^L_m(\Ml)} =\sum_{\Ml = -L}^L \Delta^L_{m \Ml} B_{\ell, L \Ml}.
\end{equation}
This is the form same as before, consistent with the general result given in Eq.~\eqref{eq:Bmintro}, where now the sum runs over $L + \Ml$ an odd integer.

\section{Flat-sky (high-momentum) limit}
\label{flat}

  \begin{figure}
    \centering
    \includegraphics[width=\columnwidth]{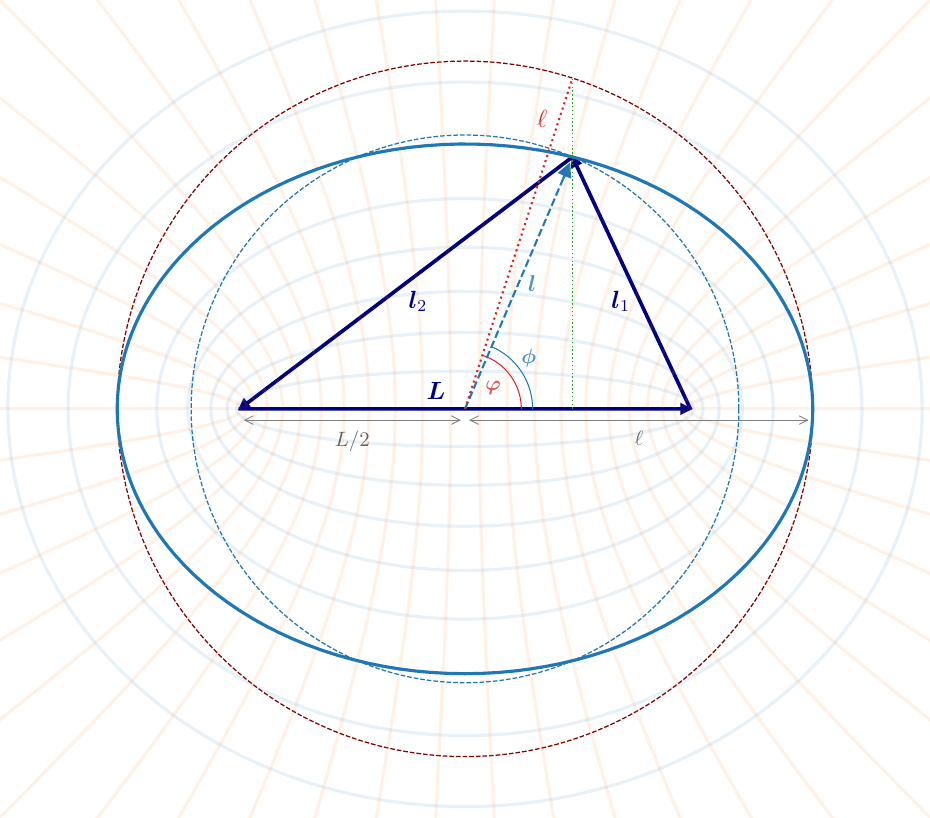}
    \caption{A flat-sky bispectrum triangle $\vL+\vl_1 + \vl_2 = 0$, with side lengths $L$, $\ell_1, \ell_2$. Scanning all possible bispectra with fixed $\ell = (\ell_1 + \ell_2)/2$ (the `short' mode in this work) and $L$ (the long mode) is the gardener's method of drawing an ellipse (in thick light blue). 
    The curved-sky expansion we introduce in this work corresponds in the flat-sky limit to a multipole (Chebyshev or $e^{im\varphi}$) expansion in $\varphi$ for triangles lying on the ellipse, where $\cos \varphi \equiv \Ml / L$, with $\Ml \equiv \ell_2 - \ell_1$ (which is not obvious from the figure). The angle $\varphi$ is indicated in the figure, and is sampled around the thin outer red circle, with $\ell \cos \varphi = l\cos \phi$.
    The angle $\phi$ (the angle between $\vl$ and $\vL$, where $\vl = (\vl_1 - \vl_2) /2$), can also be used to build a different multipole expansion of flat-sky bispectra lying on the inner thin blue circle with fixed $l$ and $L$. 
  The difference between the two angles is quadratic in the ellipse eccentricity $\epsilon = L /2\ell$, always very small for squeezed bispectra.
    The faint background lines show the lines of constant elliptic coordinates at fixed $L$, with orange hyperbolic lines at fixed $\Ml$ (and hence $\varphi$ coordinate), and blue lines at fixed $\ell$ (and hence $\mu= \cosh^{-1}(2\ell/L)$ coordinate) defining a set of confocal ellipses.
    }
    \label{fig:ellipse}
\end{figure}

In the flat-sky limit, where $L,\ell_1, \ell_2 \gg 1$, the 2D flat-sky wavevectors form a triangle with $\vL+\vl_1+\vl_2=0$.
What is the angle $\cos \varphi = \Ml/L$ in this case? 
For fixed $\vL$ and $\ell\equiv (\ell_1+\ell_2)/2$, the vertices of possible triangle shapes define an ellipse. Taking  $\vL$ to be the vector between the foci, this follows because the sum of the distances of any point on an ellipse to the two foci is a constant ($2\ell$).
However, the angle $\varphi$ is not the angular coordinate of the vertex from the centre of the ellipse. 
Instead, $\mu= \cosh^{-1}(2\ell/L)$ and $\nu=\varphi$ define the standard elliptic coordinate system\footnote{See, for example,  \href{https://en.wikipedia.org/wiki/Elliptic_coordinate_system}{Wikipedia}}, with the angle $\varphi$ being the angle shown in Fig.~\ref{fig:ellipse}. 

The flat-sky decomposition of Refs.~\cite{Lewis:2011au,Pearson:2012ba,Schaan:2018tup} instead define multipoles using the angle $\phi$ between $\vL$ and $\vl = (\vl_1 - \vl_2) /2$ at fixed $l\equiv |\vl|$, as indicated in Figs.~\ref{fig:triangles}, ~\ref{fig:ellipse}. This angle is related by $l\cos \phi = \ell \cos\varphi$ and 
\begin{equation}
	\cos \phi = \frac{\cos \varphi }{\sqrt{1 - \left(\frac{L}{2\ell} \right)^2 \sin^2\varphi}}.
\end{equation}
In the squeezed limit, $L \ll \ell_1, \ell_2$, the two foci converge on the origin, and the angles are equivalent up to quadratic order in $L/(2\ell)$.
Out of the squeezed limit, the angles and multipoles are distinct, with the Chebyshev expansion in $\varphi$ at fixed $L$, $\ell$ describing the angular dependence of triangles defining an ellipse at fixed total momentum $L+2\ell$. The symmetries in $\varphi$ are however exactly the same as described for $\phi$ in Fig.~\eqref{fig:triangles}, as they also are on the full sky.

In the flat sky limit
\begin{multline}
\sum_{2\ell,\Ml} \frac{(2L+ 1) (2\ell_1 + 1)(2\ell_2 + 1)}{4\pi} \begin{pmatrix}
		L & \ell_1 & \ell_2 \\ 0 & 0 & 0
	\end{pmatrix}^2 \\
  \rightarrow \frac{L}{2\pi^2}\int \ell\d \ell \d\varphi \frac{[1-L^2\cos^2(\varphi)/(2\ell)^2]}{\sqrt{1-L^2/(2\ell)^2}},
 \label{ellipticmeasure}
\end{multline}
which is the expected area element for elliptic coordinates, and for squeezed triangles ($\ell\gg \Ml,L$) becomes the usual measure for polar coordinates. 
In general the factor $1-L^2\cos^2(\varphi)/(2\ell)^2 = \ell_1 \ell_2/\ell^2$ gives a non-uniform weighting in angle $\varphi$.
With respect to this measure, $e^{im\varphi}\ell/\sqrt{\ell_1\ell_2}$ are a set of functions that are orthogonal in $\varphi$.
Expanding in these functions with the weight function of Eq.~\eqref{ellipticmeasure} gives a definition of the parity-even flat-sky bispectrum expansion
\begin{align} \label{eq:flatskyexp}
&b_{L\ell}^{(m)} \equiv \frac{1}{2\pi}\int \d\varphi  \sqrt{1-L^2\cos^2(\varphi)/(2\ell)^2}\,\, b_{L \ell_1 \ell_2} \cos (m\varphi),\nonumber
\\
&\frac{\sqrt{\ell_1 \ell_2}}{\ell} b_{L \ell_1 \ell_2} =  \sum_{m\ge 0} b_{L\ell}^{(m)}({2 - \delta_{m0}})\cos(m\varphi). 
\end{align}
For constant $b$, we would need to expand $\ell(\ell_1 \ell_2)^{-1/2} b_{L \ell_1 \ell_2}$ to get a pure monopole, consistent with Eq.~\eqref{ps_scale}.
In terms of $B^{(m)}_{L\ell}$ of the introduction and Eq.~\eqref{eq:Bexp}, we have
\be
B^{(m)}_{L\ell} \rightarrow \frac{2\sqrt{L\ell}}{\sqrt{\pi}[1-L^2/(2\ell)^2]^{1/4}} b_{L\ell}^{(m)}.
\ee 
In the parity odd case, the flat sky bispectrum can equivalently be expanded in $\sin(m\varphi)$ instead.

An alternative would be to expand the flat-sky bispectrum into Mathieu functions, which form a separable basis of the elliptical Helmholtz problem. However, the angular functions are not orthogonal when integrated only over $\varphi$ with the measure in Eq.~\ref{ellipticmeasure}, so a full two-dimensional mode expansion would be required.

\section{Expansion of quadratic estimators}\label{sec:QE}
   \begin{figure}
    \centering 
    \includegraphics[width=0.999\columnwidth]{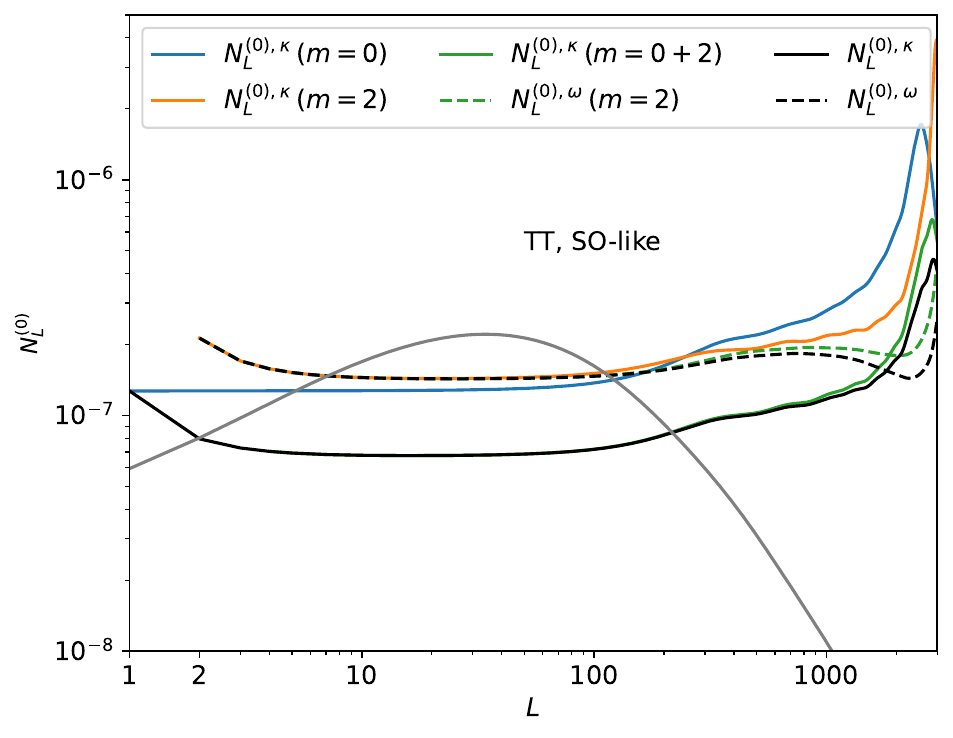}
     \includegraphics[width=\columnwidth]{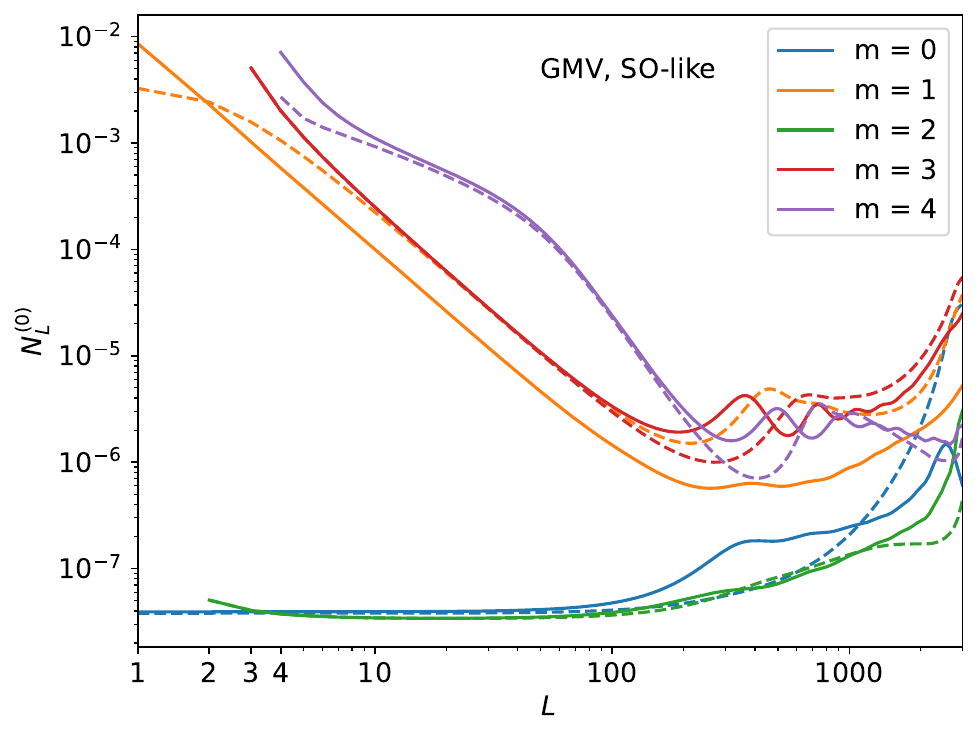}
    \caption{\emph{Top panel:} Full-sky reconstruction noise levels $N_L^{(0)}$ of the first two  moments of the temperature lensing quadratic estimators ($m=0$, probing magnification-like signatures, and $m=2$, probing shear-like signatures), computed from the orthogonal expansion given in the text. Solid lines show the case of the parity-even lensing gradient, and dashed the parity-odd lensing curl (for which there is no $m=0$ moment). This is for a configuration approaching that planned for the baseline Simons Observatory~\cite{SimonsObservatory:2018koc}. The black lines give the reconstruction noise of the standard quadratic estimators for comparison. For temperature reconstruction, the lensing curl estimator has no rotation-like contribution, and the noise of its shear part is the same at low $L$ as the lensing gradient shear. Grey shows the lensing potential power spectrum in the fiducial cosmology.
    \emph{Bottom panel:} Reconstruction noise levels for the global minimum variance (GMV) lensing gradient (solid) and curl (dashed) estimator multipole expansion, for $m=0$ to $4$, for the same experimental configuration. Odd multipoles are sourced by the anti-symmetric parts of $TE$, $TB$ and $EB$. See also Fig.~\ref{fig:GMVn0s} for more GMV $N^{(0)}$'s.}
    \label{fig:N0s}
\end{figure}
 Quadratic estimators are a family of convenient estimators designed to capture small anisotropic signals that affect underlying Gaussian isotropic fields~\cite{Hanson:2009gu}. A major application is to the gravitational lensing of the CMB~\cite{Okamoto:2003zw, Hu:2001kj}, the leading non-linear effect on the microwave background~\cite{Lewis:2006fu}. CMB lensing is now detected at more than 40$\sigma$ with \emph{Planck} and ACT~\cite{Carron:2022eyg, ACT:2023dou} using quadratic estimator techniques.
 
On large scales, a quadratic estimator works by combining two inverse-variance filtered maps with weights designed to capture the local change in the power spectrum caused by a long mode $L$ of the anisotropy source. The information on these long modes typically comes predominantly from the smallest resolved scales $\ell_1$, $\ell_2$ of the experiment (where most of the modes are), making approximately squeezed triangles the most relevant shapes.

In general, an un-normalized quadratic estimator using two observed fields $X$ and $Z$ can be written in terms of their inverse-variance weighted versions (denoted with a bar)
\begin{align}\label{eq:QEdef}
	\bar x_{LM}[\bar X, \bar Z] = \frac{(-1)^M}2 \sum_{\ell_1 m_1, \ell_2 m_2}& \begin{pmatrix}
		L & \ell_1 & \ell_2 \\ -M & m_1 & m_2 
	\end{pmatrix} \\
	&\times W^x_{L \ell_1 \ell_2} \bar X_{\ell_1 m_1}\bar Z_{\ell_2 m_2}. \nonumber
\end{align}
As reviewed in Appendix~\ref{app:QEs}, optimal weights $W^x$ are given by the response bispectrum (Eq.~\ref{lenskernel}) in a fiducial model:
\begin{equation}
W^x_{L \ell_1 \ell_2} =\left[f^{x XZ }_{L \ell_1 \ell_2} \right]^*.
\label{QEweightbi}
\end{equation} 
\subsection{Single field}
Consider first for simplicity the case where the two fields are equal. Neglecting real-world non-idealities, for optimally-filtered estimators the spectrum of the map $\bar X$ is given by $\bar C_\ell = 1/ C_\ell^{\rm tot}$, where $C_\ell^{\rm tot}$ is the total power of $X$.
The leading covariance term (usually called `$N_L^{(0)}$', after normalization by the estimator response to the anisotropy source) is given by
\begin{equation}
	\av{\bar x_{LM}\bar x_{LM}^*} = \frac1 {2 (2L + 1)} \sum_{\ell_1 \ell_2}  \frac{|W^x_{L \ell_1 \ell_2}|^2}{C^{\rm tot}_{\ell_1} C^{\rm tot}_{\ell_2}}.
\end{equation}
This is essentially the same as Eq.~\eqref{eq:FLB}, so we can proceed in the same manner.
If we expand the weights, replacing $B_{L\ell_1 \ell_2} \rightarrow  w_{L\ell_1\ell_2}\equiv W_{L\ell_1 \ell_2}/ \sqrt{C^{\rm tot}_{\ell_1} C^{\rm tot}_{\ell_2} }$ in Eqs.~\eqref{eq:Bexp} and \eqref{eq:Bexpodd}, we obtain
a family of pairwise-orthogonal estimators $\bar x^{(m)}_{LM}$, with diagonal noise
\begin{equation}\label{eq:n0}
	\av{\bar x^{(m)}_{LM}\bar x^{(m')*}_{LM}} =  \delta_{mm'}\frac {2 - \delta_{m0} }{2(2L + 1)} \sum_\ell |w^{(m)}_{L\ell}|^2.
\end{equation}
The estimators are explicitly given by
\begin{align} \nonumber
	&\bar x_{LM}^{(m)}[\bar X, \bar X] = \frac{(-1)^M}2 \sum_{\ell_1 m_1, \ell_2 m_2} \begin{pmatrix}
		L & \ell_1 & \ell_2 \\ -M & m_1 & m_2 
	\end{pmatrix} \\ \label{eq:QEm}
	&\times(-1)^{\lfloor{(L-\Ml)/2}\rfloor}\Delta^L_{m \Ml}(2 - \delta_{m0})\: w^{(m)}_{L\ell} \frac{\bar X_{\ell_1 m_1}}{\sqrt{\bar C_{\ell_1}}}\frac{\bar X_{\ell_2 m_2}}{\sqrt{\bar C_{\ell_2}}}.
\end{align}
For these estimators the unnormalized variance is also always equal to their responses $\mathcal R_L$ to the signal, and hence also to the inverse normalized estimator noise $N^{(0)}_L$.

The noise curves for the temperature-only lensing gradient mode (labelled $\kappa$) and curl (labelled by the field rotation $\omega$) estimators are shown on Fig.~\ref{fig:N0s}. 
The monopole and quadrupole results are well described at low-$L$'s by the full-sky version of the well-known 
squeezed limits associated to the information from local changes of the power spectrum under the action of large-scale magnifying $(m=0)$ or shearing lenses $(m=2)$~\cite{Bucher:2010iv,Schmittfull:2013uea,Pearson:2012ba}, 
\begin{align} \nonumber \label{eq:RL}
	\mathcal R_L^{\kappa(m=0)} &\sim \frac 12 \sum_\ell \left ( \frac{2\ell + 1} {4\pi} \right) \left(\frac{\Cgrad_\ell}{C_\ell^{\rm tot}} \right)^2\left(\frac{\d \ln \ell(\ell+1)\Cgrad_\ell}{\d \ln (2\ell+1)} \right)^2 \\
		\mathcal R_L^{\kappa(m=2)} &\sim \frac{(L-1)(L + 2)}{L(L + 1)} \\ &\times\frac 12 \sum_\ell \left ( \frac{2\ell + 1} {4\pi} \right)  \left(\frac{\Cgrad_\ell}{C_\ell^{\rm tot}} \right)^2\frac 12\left(\frac{\d \ln \Cgrad_\ell}{\d \ln (2\ell+1)} \right)^2 \nonumber \\
		\mathcal R_L^{\omega(m=2)} &\sim \mathcal R_L^{\kappa(m=2)}, \nonumber
\end{align}
where $\Cgrad_\ell \equiv C^{T\nabla T}_\ell$ is the non-perturbative gradient response\footnote{Eq.~\eqref{eq:RL} holds for the optimally weighted estimator using this non-perturbative spectrum $\tilde C_\ell$. More generally, in the squared numerators of this equation, one of the two spectra is the one used for the estimator weighting, and the other this non-perturbative spectra.} spectrum~\cite{Lewis:2011fk}, approximated here as a continuous function at high $\ell$.
These equations follow from the temperature lensing weights by expanding $\ell_1$ and $\ell_2$ around $\ell$ to $\mathcal{O}(L^2/\ell^2)$, and computing the resulting $\Ml$ moments.

The temperature curl mode estimator response is pure shear, and equal to the shear part of that of $\kappa$. This is because a locally constant pure rotation $\omega$ has no observable effect on the local temperature power spectrum, but $\omega$ enters the $B$-mode of the shear-field in the same manner than $\kappa$ enters its $E$-mode. The $L$-dependent prefactor (absent in flat-sky calculations) originates from the relation between the convergence and shear fields on the curved-sky (see appendix \ref{app:sqz} for a detailed discussion), and makes a substantial difference at the very lowest lensing multipoles. The situation is different and opposite in the case of the polarization $EB$ (or $TB$) estimator: a locally constant magnification of the polarized Stokes parameters is unobservable, but a pure rotation is, since it induces a non-zero $EB$ spectrum; the lensing gradient mode is pure shear and equal to the shear part of the lensing curl, which also has a perfectly white rotation component.
\subsection{Multi-field minimum variance}

   \begin{figure}
    \centering
    \includegraphics[width=\columnwidth]{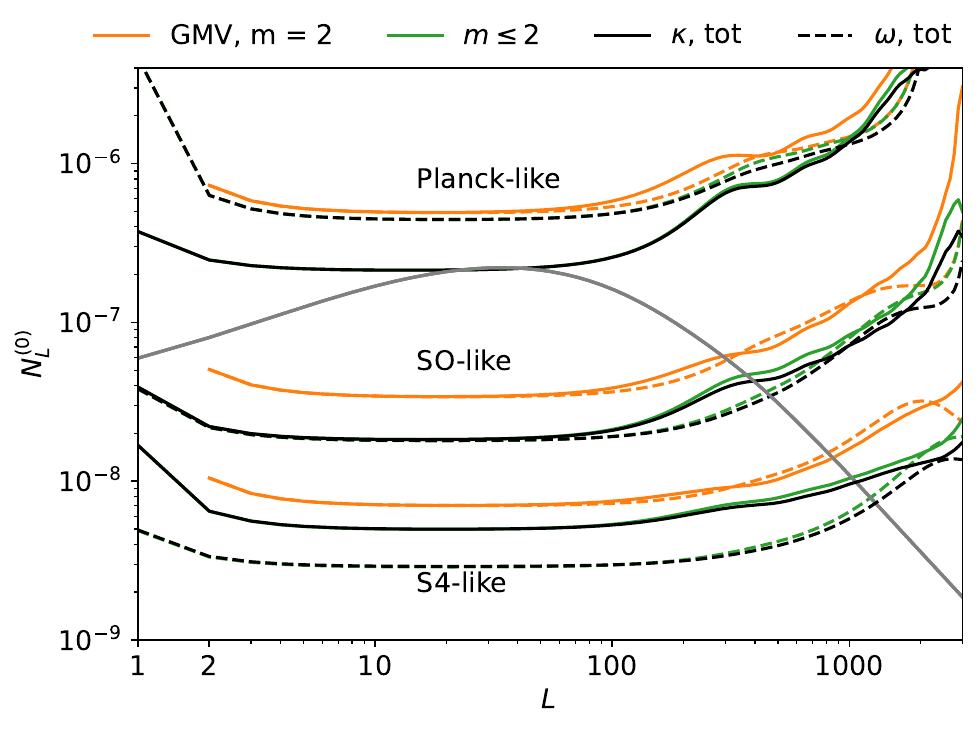}
    \caption{Full-sky reconstruction noise levels $N_L^{(0)}$ of the shear-only piece of the GMV lensing quadratic estimators ($m=2$, orange lines), for a \emph{Planck}-like, SO-like and S4-like configuration from top to bottom. Solid lines show the case of the lensing gradient ($\kappa$), and dashed the lensing curl reconstruction ($\omega$). The black lines give the reconstruction noise of the optimal quadratic estimators for comparison, and green the reconstruction noise of the first three moments combined ($m=0, 1$ and $2$). The lensing dipole is always a pure magnification-like (or pure rotation-like, in the curl case) rather than a shear-like signal, so that the value of the black curve at $L=1$ always indicates the white noise level of the $m=0$ piece. The shear-pieces of the gradient and curl mode noise are always equal at low lensing multipoles. With the optimal quadratic estimators (the black lines), the lensing gradient mode has lower noise than the curl mode for \emph{Planck}-like experiments, but higher noise for S4-like experiments. This is because of domination by the parity-even $\kappa TT$ and parity-odd $\omega TT$ bispectra at high noise levels, but conversely by the parity-odd $\kappa EB$ and parity-even $\omega EB$ at low noise polarization noise levels. Hence, for deep polarization experiments, and for all other things being equal, the shear-only piece becomes suboptimal (by a factor of $3$) for the curl mode, but gets better than the $m=0$ piece and generally closer to optimality for the lensing gradient (in this S4 configuration with $\ell_{\rm max} = 4000$ there is non-negligible information coming from temperature, so that the orange solid line is visibly higher than the black one). The transition happens at the SO-like configuration, which has approximately equal shear and convergence, lensing and curl low-$L$ reconstruction noise. The grey curve is the fiducial lensing gradient power spectrum for reference (the post-Born lensing curl spectrum is much smaller and not shown).
    \label{fig:GMVn0s}}
\end{figure}
This leads us to consider the combination of estimators in more detail.
Let $\clmat$ (as in appendix~\ref{app:QEs}) be the covariance of a set of fields (inclusive of noise). Optimal inverse-variance filtering the set of fields gives $\bar X = \sum_Z \left[\clmat^{-1}\right]^{XZ} Z$, and hence $\bar{\clmat} = \clmat^{-1}$. The optimal quadratic estimator built from these fields is sometimes called the Global  Minimum Variance (`GMV')~\cite{Maniyar:2021msb} estimator. In analogy with \eqref{eq:QEm} we can write its multipole expansion as
\newcommand{\Xvec}[1]{\boldsymbol{#1}}
\begin{align} \nonumber
	&	\bar x_{LM}^{\text{GMV}(m)} \equiv \frac{(-1)^M}{2}\sum_{\ell_1 m_1,\ell_2 m_2} \begin{pmatrix}
			L & \ell_1 & \ell_2 \\ -M  & m_1 & m_2
		\end{pmatrix} \\ \nonumber &\times  (2 - \delta_{m0}) \sum_{X Z} w^{(m), XZ}_{L\ell}\left[\bar{\clmat}^{-1/2} \bar {\Xvec Y}\right]_{\ell_1 m_1}^{X}\left[\bar{\clmat}^{-1/2} \bar {\Xvec  Y }\right]^{Z}_{\ell_2 m_2} \\ &\times (-1)^{\lfloor{(L-\Ml)/2}\rfloor} \Delta^L_{m \Ml} 
\label{EQpol}
\end{align}

with resulting Gaussian noise
\begin{equation}\label{eq:n0GMV}
	\av{\bar x^{\text{GMV}(m)}_{LM}\bar x^{\text{GMV}(m')*}_{LM}} =   \delta_{mm'}\frac {2 - \delta_{m0} }{2(2L + 1)} \sum_{\ell, XY}\left|w^{(m), XY}_{L\ell}\right|^2.
\end{equation}
Explicitly, the weights $w^{(m), XY}_{L\ell}$ are obtained using
\begin{align}
 w^{XY}_{L\ell_1\ell_2}\equiv 
\sum_{Z_1 Z_2} \bar{\clmat}_{\ell_1}^{1/2, XZ_1}\bar{\clmat}_{\ell_2}^{1/2, YZ_2} W^{x Z_1Z_2}_{L \ell_1 \ell_2}
\end{align}
for $B$ in Eqs.~\eqref{eq:Bexp} and \eqref{eq:Bexpodd}. The most relevant cases of interest being that of reconstruction from polarization ($X,Y \in (E, B))$, which combines the $EE, EB$ and $BB$ pairs, and in combination with temperature $(X, Y \in (T, E, B))$ which uses all of them. Reconstruction noise curves in this later case are shown in Fig.~\ref{fig:GMVn0s}. For deep experiments like CMB-S4, $EB$ typically dominates, with low-$L$ behaviour to $O(L/\ell)^2$ corrections given by (dropping small $\tilde C_\ell^{BB}$ terms, and neglecting $C_\ell^{TE}$ in the filtering matrix, see appendix~\ref{app:sqz})
\begin{align} \label{eq:EBsqdz}
	\mathcal R_L^{ \kappa(m=0)} &\sim 0 \quad \text{           $(EB)$} \\
		\mathcal R_L^{\omega(m=0)}&\sim 4 \sum_\ell \left ( \frac{2\ell + 1} {4\pi} \right) \left(\frac{\tilde C^{EE}_\ell}{C_\ell^{EE,\rm tot}} \right)\left(\frac{\tilde C^{EE}_\ell}{C_\ell^{BB,\rm tot}} \right) \nonumber \\ \nonumber
	\mathcal R_L^{ \kappa(m=2)} &\sim  \mathcal R_L^{\omega(m=0)}\frac12 \frac{(L-1)(L + 2)}{L(L + 1)} \sim \mathcal R_L^{ \omega(m=2)} \nonumber.
\end{align}
More squeezed limits are given in appendix~\ref{app:sqz}.

   \begin{figure}
    \centering
    \includegraphics[width=0.999\columnwidth]{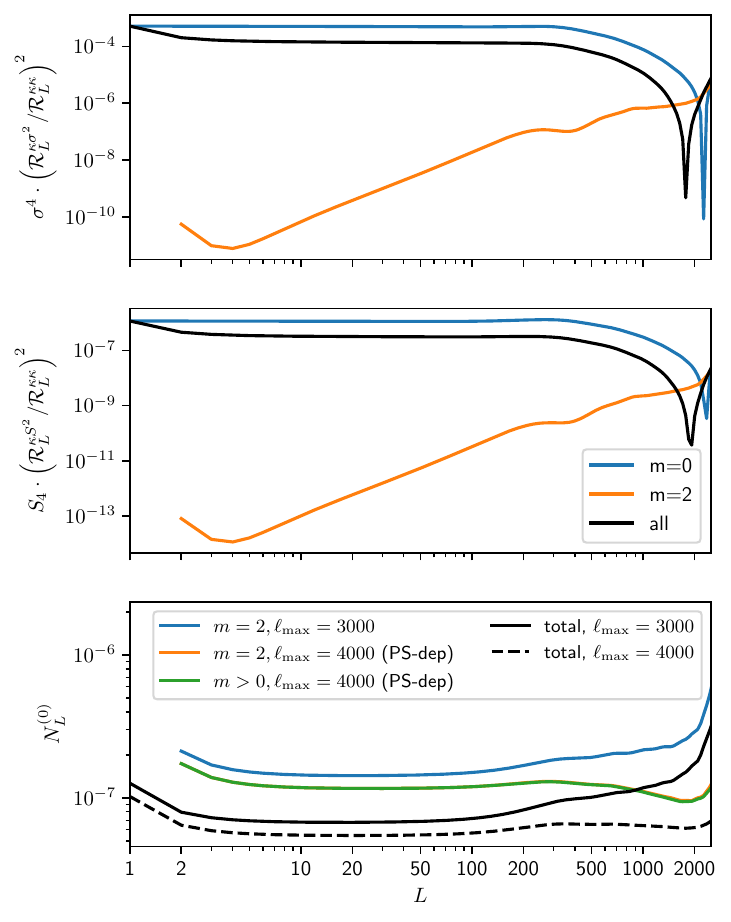}
    \caption{Response of the $TT$ lensing reconstruction spectrum to noise inhomogeneities (upper panel) and point source contamination (middle panel). These are to a good approximation monopolar (blue) and the $m=2$ multipole estimator (orange) are essentially free of these systematics at low lensing multipoles. See text for the normalization of these curves. This is for a Simons-Observatory-like configuration. Alternatively, one can exactly project the contamination on all scales, and try and extend the range of CMB multipoles used for the lensing reconstruction. This is shown on the bottom panel in the case of point sources. Over these scales, the monopole-projected estimator ($m>0$) is only very slightly better than using $m=2$ alone.\label{fig:PS}  
    }
  \end{figure}

  \subsection{Robustness}
In practice, quadratic estimators are prone to contamination by various sources, at high and low lensing multipoles. 
As an example, we consider contamination from unclustered sources with radially-symmetric profiles, which includes as a special case Poisson point sources and uncorrelated but anisotropic noise.
We focus on the case of temperature quadratic reconstruction, which is the most relevant, since in polarization $EB$ typically dominates the relevant lensing gradient signal but already has much lower foreground contamination, and is already mostly shear-only.

We assume random unclustered sources with circularly symmetric profiles $I(\cos\theta)$, so that
the induced beam-deconvolved temperature correlation is
\begin{equation}
\la T^{(S)}(\hn_1)T^{(S)}(\hn_2)\ra = \int d\hn~ I(\hn_1\cdot \hn)\:I(\hn_2\cdot \hn)~S^2(\hn),
\end{equation}
where $S^2(\hn)$ measures the spatially-varying source density.
Expanding $I(\cos\theta)$ into Legendre modes $I_\ell$ and using the spherical harmonic addition theorem,
the beam-deconvolved CMB mode covariance then has a contribution
	\begin{multline}
\av{T^{\rm obs}_{\ell_1 m_1}T^{\rm obs}_{\ell_2 m_2}}\ni I_{\ell_1} I_{\ell_2}  S^2_{LM}(-1)^M 	 \begin{pmatrix}
	L & \ell_1 & \ell_2 \\ -M & m_1 & m_2
\end{pmatrix}\\\times \begin{pmatrix}
	L & \ell_1 & \ell_2 \\ 0 & 0 & 0 
\end{pmatrix}\sqrt\frac{(2\ell_1 + 1)(2\ell_2 + 1)(2L + 1)}{4\pi}.
\label{eq:psvar}
\end{multline}

This anisotropic variance map will induce a response in the  gradient part of the idealized $TT$ estimator of Eq.~\eqref{eq:QEdef},
\begin{equation}
	\bar \kappa_{LM} \sim \mathcal R^{\kappa\kappa}_L \kappa_{LM} + \mathcal R^{\kappa S^2}_L S^2_{LM}   + \cdots 
\end{equation}
where $S^2_{LM}$ are the harmonic coefficient of the variance map in the relevant units, and we introduced $\mathcal R_L^{\kappa S^2}$ to characterize the response.
In particular, the response of the $\kappa^{(m)}$ estimator (Eq.~\ref{eq:QEm}) will be proportional to
\begin{equation}
	\mathcal R_L^{\kappa^{(m)} S^2} \propto \avM{\mathcal T_m^L(\Ml) \frac{h_{\ell_1}}{\sqrt{\bar{C}_{\ell_1}}} \frac{h_{\ell_2}}{\sqrt{\bar{C}_{\ell_2}} }}
\label{response_h}
\end{equation}
with
\begin{equation}
	h_\ell = I_\ell \sqrt{\frac{(2\ell + 1)}{(4\pi)^{1/2}}}.
\label{h_ell_ps}
\end{equation}

Since the foregrounds are most important at high $\ell$, and for low $L$ most of the lensing signal is in squeezed shapes, much of the signal will be with $L\ll \ell$.
Also $h_\ell$ and $\bar{C}_\ell$ are smooth functions, $h_{\ell_1}h_{\ell_2} \approx h_\ell^2 + \mathcal{O}(L^2/\ell^2)$ for $L\ll \ell$, so the source signal is mostly monopole ($\mathcal R_L^{\kappa^{(m)} S^2} \propto \delta_{m0}$) with the quadrupole suppressed by $(L/\ell)^2$ for $L\ll 200$ where expanding $\bar{C}_\ell$ is also accurate. Projecting the monopole, or using just the quadrupole, can therefore remove the bulk of the contaminating signal at low $L$.

For comparison, the response of the full lensing estimator to contamination from Eq.~\eqref{eq:psvar} can be calculated from Eq.~\eqref{eq:d1}, giving
\begin{align}
	\frac 12 \sqrt{L(L + 1)}&\mathcal R^{\kappa S^2}_L =- 2\pi \int_{-1}^1 d\mu\: d^L_{-10}(\mu) \bigg \{ \\
&\nonumber \left (\sum_\ell \left( \frac{2\ell + 1}{4\pi}\right) \frac{1}{C_\ell^{\rm tot}}  I_\ell d_{00}^L(\mu)\right) \\
&\!\!\!\!\!\!\left.\times \left (\sum_\ell \left( \frac{2\ell + 1}{4\pi}\right) \frac{\Cgrad_\ell}{C_\ell^{\rm tot}}  I_\ell  \sqrt{\ell(\ell + 1)} d_{10}^L(\mu)\right) \right\}. \nonumber
\end{align}

As a concrete example, the contribution of a point-source bias ($I_l=1$)
\begin{equation}
		S_4\left( \frac{\mathcal R^{\kappa^{(m)} S^2}_L}{\mathcal R_L^{\kappa^{(m)} \kappa}} \right)^2,
\end{equation}
is shown on the lower panel of Figure~\ref{fig:PS}, with an arbitrary normalization 
\begin{equation}
S_4 = 3e^{-13}\mu \rm{K}^4,
\end{equation}
as measured on a \emph{Planck} map after component separation~\cite{Aghanim:2018oex} using a measurement of the trispectrum. The contamination is substantially suppressed on large scales, though less efficiently on small scales.

At low lensing multipoles, masking, and unavoidable inhomogeneities in the noise maps sourced by the scan strategy typically induces very strong noise anisotropy leading to a large mean field signals~\cite[e.g., App.~B has a detailed discussion of the \emph{Planck} mean field]{Aghanim:2018oex}. Spatially-varying uncorrelated noise is equivalent to point sources, but with the noise added on the beam-convolved map, so that $I_\ell =b_\ell^{-1}$ where $b_\ell$ are the moments of a circularly-symmetric beam.
In the upper panel of Fig.~\ref{fig:PS} we show  the contributions of these responses to the lensing convergence power spectrum,
\begin{equation}
	\sigma^4 \left( \frac{\mathcal R^{\kappa^{(m)} \sigma^2}_L}{\mathcal R_L^{\kappa^{(m)} \kappa}} \right)^2,
\end{equation}
for an arbitrary value of $\sigma = 7 \mu \rm{K arcmin}$ (crudely comparable to the value found on component separated \emph{Planck} noise maps at $L \sim 10$), and a Gaussian beam with FWHM 3 arcmin.
The corresponding induced contribution to the convergence spectrum is shown as the black line in the figure for comparison.

Noise inhomogeneities do not affect the temperature curl-mode reconstruction. They could potentially affect the curl in polarization, where $EB$ dominates. In polarization, there are two types of variance maps, the spin-0 $\av{_{\pm 2}P\:_{\pm 2}P^*} =  \av{Q^2} + \av{U^2}$, and the spin-4 $\av{_{\pm 2}P \:_{\mp 2}P^*}=\av{Q^2} - \av{U^2} \pm 2i \av{QU}$.
Again, from parity arguments, the curl estimator can only couple of the curl mode of the spin-4 map, and not to the spin-0 map which is gradient only. The spin-4 component is typically very small for CMB experiments, except possibly in places which are very sparsely scanned and the signal hence very weak, so we do not consider this further here.
  
    It is also possible to make the response exactly immune to sources (or noise anisotropy) on all scales by redefining the bispectrum to include the factors of $h_\ell$ (Eq.~\ref{h_ell_ps}) in the weights that are multipole expanded\footnote{Including $h_\ell$ in the weights means the $T_{\ell m}$ are divided by $h_\ell$ in the estimator, so the response of the estimator to point sources becomes pure monopole consistent with Eq.~\eqref{ps_scale}.}, rather than $\sqrt{\bar{C}_\ell} =1/\sqrt{C^{\rm tot}_\ell}$ used in Eq.~\eqref{eq:QEm}. In this case, the estimators are no longer perfectly orthogonal to each other, but instead exactly project the source signal for all $m>0$ (up to band limit effects, see Appendix~\ref{bandlimits}). 
    For most plausible source profiles $h_\ell$ is very smooth in $\ell$, and there is no longer an oscillatory part from $\bar{C}_\ell$. This means that the squeezed approximation should now hold over a wider range of $L\ll \ell$, so this estimator should also be robust to profile misestimation to larger $L$\footnote{$r_{\ell_1}r_{\ell_2} \approx r_\ell^2 + \mathcal{O}(L^2/\ell^2)$ where $r_\ell\equiv I_\ell/I^{(\rm fid)}_\ell$ is the ratio between the true and fiducial assumed source profile.} as well as exactly projecting sources with the assumed mean profile.
    Since the estimator is more robust, one can potentially then use a larger range of CMB multipoles. We test this on the lower panel of Fig.~\ref{fig:PS} for the case of point sources, comparing the noise curves at 3000 and 4000, without and with PS-deprojection. Substantial gain is seen at high-$L$ compared to the full standard estimator noise (shown in black). 

 Our multipole estimators can therefore be used to project out classes of possible contaminating signals that are mostly monopolar. In practice, for point sources one could account for the average profile using a general $I_\ell$ (rather than $I_\ell=1$ as assumed for the specific examples above). The monopole-projected or quadrupole estimators would then be immune to these sources as long as the average profile is well-enough known, and also robust to mis-modelled profiles at low $L$ where the squeezed limit holds. The cost is an increase in noise, making the method noisier than projecting a full specific shape (profile hardening,~\cite{Namikawa:2012pe,Sailer:2020lal}) while being more robust, as for the flat-sky estimators~\cite{Schaan:2018tup}.

\section{Conclusions}

We presented a decomposition of spherical bispectra and quadratic estimator kernels into a set of modes with distinct dependence on the bispectrum triangle shape at fixed total momentum. With suitable choice of weights, the set of modes are orthogonal. In squeezed shapes they are dominated by the lowest moments which encapsulate the leading monopolar and quadrupolar shape dependence (magnification and shear in the case of lensing). Alternatively, the weights can be chosen to exactly project specific classes of possible contaminating signals. Consistency of data constraints between different components of the moment expansion gives a powerful consistency check on contamination by qualitatively distinct possible contaminating signals. We showed explicitly that point sources and noise mean fields appear mostly in the monopole, so a monopole projection is one way to make an estimator that is insensitive to these signals (at the expense of increasing the noise compared to a more specific signal projection, e.g. using bias hardening~\cite{Namikawa:2012pe}).
In specific cases the moments themselves may also largely isolate distinct physical contributions to the signal of interest.

In terms of mixed-field bispectra with specific parity, the decomposition is equivalent to expanding into a set of orthogonal polynomials that are a quantized version of the Chebyshev polynomials of type 1 (for even parity), or type 2 (for odd parity). These polynomials can be calculated easily from recursion relations, and have a simple analytic form related to a configuration-space rotation of the full (unreduced) bispectrum.
Although the corresponding estimators are not directly separable in real space, the leading moments of most interest in many cases can still be evaluated efficiently. For squeezed bispectrum shapes, in the flat-sky limit the expansion is equivalent to previous angular expansions used on the flat sky. However, for general shapes the high-momentum limit gives a distinct flat-sky mode expansion that differs from previous results away from the squeezed limit. 
The modes presented here also allow a consistent treatment of the largest angular scales.

\begin{acknowledgements}
We thank Frank Qu Zhu for useful comments on a draft.
AL was supported by the UK STFC grant ST/X001040/1. 
JC acknowledges support from a SNSF Eccellenza Professorial Fellowship (No. 186879).
\end{acknowledgements}

\appendix

\begin{widetext}

\section{Multi-field multipole expansion}\label{ap:multifield}
When there are multiple fields we can define configuration space multipoles that have the symmetry of spherical harmonics using
\be
B^{ij}_{\lplus,L\Ml } \equiv (-1)^{\lfloor{(L-\Ml )/2}\rfloor} \left(B^{(ij)}_{L (\lplus-\Ml /2) (\lplus+\Ml /2)} + iB^{[ij]}_{L (\lplus-\Ml /2) (\lplus+\Ml /2)}\right).
\label{Bijexpand}
\ee
This can be expanded into multipoles $B_{\ell L}^{ij(m)}$ following Eq.~\eqref{eq:Bmintro},
where the antisymmetric part can give rise to odd $m$ (with $L+\Ml $ still even for the parity-even case).
The inverse is
\be
B^{ij}_{L (\lplus-\Ml /2) (\lplus+\Ml /2)}  \equiv (-1)^{\lfloor{(L-\Ml )/2}\rfloor}
\sum_{m\ge 0} (2 - \delta_{m0}) \Delta^L_{m\Ml }
\left(B^{(ij) (m)}_{\lplus L } - iB^{[ij] (m)}_{\lplus L}  \right).
\ee
The symmetric part only contributes with $m$ even, the anti-symmetric part gives odd $m$.
For even $m$ the real part of $B^{ij(m)}$ is parity even, and the imaginary part parity odd.
For odd $m$ the imaginary part of $B^{ij(m)}$ is parity even, and the real part parity odd.
In terms of the quadratic estimator of Eq.~\eqref{EQpol}, the $W^{x,XZ}_{L\ell_1 \ell_2}$ can each also be expanded directly into $\T$ or $\U$ (depending on the parity of the pair) for all odd and even multipoles $m$, with the $i$ factors of Eq.~\eqref{Bijexpand} cancelling in the final form of the quadratic estimator.
\section{Quantized Chebyshev polynomials} \label{app:poly}

In this appendix we give a more complete description of the discrete polynomials that are pairwise orthogonal when summed over $\Ml$ with weight $w_{L\Ml}$.
We define
\begin{equation}
	\T^L_m(\Ml) \equiv \frac{\dlp L m \Ml}{\dlp L 0 \Ml}
	\quad w_{L\Ml} \equiv \left(\dlp L 0 \Ml\right)^2  \quad (\text{for }L + \Ml \text{ even})
\end{equation}
and
\begin{equation}
	\U^L_m(\Ml) \equiv \frac{\dlp L m \Ml}{\dlp L 1 \Ml}
	\quad w_{L\Ml} \equiv \left(\dlp L 1 \Ml \right)^2  \quad (\text{for } L + \Ml \text{ odd}).
\end{equation}
Here $\dlp L m n \equiv d^L_{mn}(\pi /2)$ are elements of the small Wigner $d$-matrix that rotate by $\pi/2$ and have the symmetries
\be
 \Delta^L_{mm'}=(-1)^{L+m}\Delta^L_{m,-m'} =
(-1)^{m-m'}\Delta^L_{m'm} = (-1)^{L-m'}\Delta^L_{-mm'}.
\ee
They also obey the orthogonality relations $\sum_\Ml \Delta^L_{m\Ml}\Delta^L_{m'\Ml} = \delta_{mm'}$ and $\sum_\Ml \Delta^L_{m\Ml}\Delta^L_{\Ml m'} = d^L_{mm'}(\pi) = (-1)^{L+m}\delta_{m,-m'}$.

Seen as a function of $\Ml$, $\T^L_m$ is a polynomial of degree $m$ in $\Ml$, and $\U^L_m$ of degree $m-1$ in $\Ml$: this follows from the three-term recursion relation~\cite[p.93]{AngularMom}
\begin{equation}
	2\Ml \dlp Lm\Ml = \sqrt{(L + m)(L  -m + 1)}\dlp L {m-1}\Ml + \sqrt{(L - m)(L + m + 1)}\dlp L {m+1}\Ml,
\end{equation}
with initial polynomials  $\T^L_0(\Ml) = 1$, $\T^L_1(\Ml) = \Ml /\sqrt{L(L+1)}$, and  $\U^L_0(\Ml) = 0, \U^L_1(\Ml) = 1$. The polynomials are orthogonal with respect to the weights $w_{LM}$ as defined above: we have 
\begin{equation}\label{eq:ortho}
\begin{split}
	\sum_{\Ml, (L + \Ml)\: \text{even}} \:w_{L\Ml} \T^L_m(\Ml)\T^L_{m'}(\Ml) &=\delta_{mm'} \left\{ \begin{matrix}
		\frac 12 & (m > 0) \\ 1 & (m = 0)		
	\end{matrix} \right.\\
	\sum_{\Ml, (L + \Ml)\: \text{odd}} \:w_{L\Ml}\: \U^L_m(\Ml)\U^L_{m'}(\Ml) &= \delta_{mm'}\frac 12 .
\end{split}
\end{equation}
An equivalent result in terms of Wigner-$D$ matrices has been given by~\cite{Zhong:2023pak}, and can be seen for example using the relation~\cite{Risbo:1996aa,Huffenberger:2010hh},
\begin{equation}\label{eq:Fourier}
	d^L_{mm'}(\cos \theta) =  \sum_{\Ml = -L}^L i^m \Delta^L_{\Ml m}  e^{-i\Ml\theta} \Delta^L_{\Ml m'}(-i)^{m'},
\end{equation}
which holds for any $\theta$ and is equivalent to Eq.~\eqref{xeigs} in the main text.
The left-hand side is the matrix representing a rotation by $\theta$ around the $y$-axis, and the right-hand side its decomposition into the series of rotations constructed such that the rotation by $\theta$ is made around the $z$-axis, giving a pure phase $e^{-i \Ml \theta}$. This representation provides a complete eigendecomposition of the matrix $d^L(\cos \theta)$.
 The real part of the corresponding eigenvalue equation implies that if $L+M$ is even
\ba
\sum_{M', (L+M')\text{even}} d^L_{MM'}(\theta) \left[(-1)^{(L+M')/2}\Delta^L_{mM'}\right]= \cos(m\theta) \left[(-1)^{(L+M)/2} \Delta^L_{mM}\right].
\ea
Similarly for the imaginary part with $L+M$ odd:
\ba
\sum_{M', (L+M')\text{odd}} d^L_{MM'}(\theta) \left[(-1)^{(L-1+M')/2}\Delta^L_{mM'}\right]= \cos(m\theta) \left[(-1)^{(L-1+M)/2} \Delta^L_{mM}\right],
\ea
so the eigenvectors restricted to parity subspaces are in both cases are entirely real. Eq.~\eqref{eq:ortho} then follows from the completeness of the two sets of eigenvectors, with the half factors to ensure that states that are symmetric or anti-symmetric about the equator both have norm preserved under rotation~\cite{Zhong:2023pak}.

The polynomials also obey the completeness relations
\begin{equation}
\frac 12 +  \sum_{m = 1}^L \T^L_m(\Ml)\T^L_m(\Ml') = \frac{\delta_{\Ml\Ml'}}{2w_{L\Ml}},\quad   \sum_{m = 1}^L \U^L_m(\Ml)\U^L_m(\Ml') = \frac{\delta_{\Ml \Ml'}}{2w_{L\Ml}}.
\end{equation}
The weight functions in the parity-odd and parity-even cases are related by~\cite[p113]{AngularMom}
\begin{equation}
	\dlp L 1 \Ml = -\sqrt{\frac{L^2 - \Ml^2}{L (L + 1)}} \dlp {L-1} 0 \Ml \quad (L + \Ml  \text{ odd}).
\end{equation}
With $\cos \varphi \equiv 2\Ml / (2 L + 1) = x$, the flat-sky correspondence to the Chebyshev polynomials of type 1 and 2 is as follows,
\begin{equation}
\T^L_m(\Ml) \rightarrow T_m(x) =\cos m \varphi,\quad 	\U^L_m(\Ml) \rightarrow U_{m -1}(x) = \frac{\sin m \varphi}{\sin \varphi},
\end{equation}
with the weight functions approaching the corresponding weights proportional to $(1-x^2)^{-1/2}$ and $(1- x^2)^{1/2}$ respectively.

The $\T$ and $\U$ polynomials form a quantized version of the usual Chebyshev polynomials of the first and second kind, in an analogous way that the discrete Chebyshev polynomials\footnote{The terminology is confusing, because these were also used by Chebyshev in the context of interpolation, but are not related to the normal continuous (or our quantized) Chebyshev polynomials.} $f^L_m(M)$ are a quantized version of the Legendre polynomials $P_L(\cos\theta)$~\cite{Garg22}.  These discrete polynomials are eigenvectors of $|d^L_{MM'}|^2$~\cite{Meckler58}:
\begin{equation}
\sum_{M'=-L}^{L} |d^L_{MM'}(\theta)|^2 f^L_m(M') =  P_L(\cos(\theta))f^L_m(M),
\end{equation}
which can be compared with our analogous Eq.~\ref{xeigs}.

\subsection{Other results}
From \eqref{eq:Fourier} it directly follows that
\begin{equation}
d^L_{m0} (\cos \theta) = (-i)^{m} 	\T^L_m\left(i\frac{d }{d\theta}\right) d^L_{0 0}(\cos \theta),
\end{equation}
a `curved-sky' analog to the known `flat-sky' relation between Chebyshev polynomials and Bessel functions,
\begin{equation}
	J_n(x)= i^n T_n\left(i\frac{ d}{dx}\right) J_0(x).
\end{equation}
Another correspondence is the Jacobi-Anger expansion; for $|\Ml| \le L ,\Ml + L$ even, 
\begin{align}
	e^{-i \Ml \theta} &= 	d^L_{00}(\theta) + 2 \sum_{m =1}^L d^L_{m0}(\theta) i^m \mathcal T_m^L(\Ml) = J_0(L \theta) + 2 \sum_{m=1}^\infty  J_m(L\theta) (-i)^m T_m\left( \frac \Ml  L \right).
\end{align}
On the right-hand side we have the Jacobi-Anger expansion of $e^{-i z \cos \phi}$ for $\cos \phi =  \Ml / L $ and $z = L \theta$.

More generally, the polynomials act like raising operators for even and odd Wigner-$d$:
\bea
d^L_{mm'}(\cos\theta)+(-1)^{m'} d^L_{m-m'}(\cos\theta) &=& i^{-m}\T_m\left(i\frac{d}{d\theta}\right) 2d^L_{0m'}(\cos\theta) \\
d^L_{mm'}(\cos\theta)-(-1)^{m'} d^L_{m-m'}(\cos\theta) &=& i^{1-m}\U^L_m\left(i\frac{d}{d\theta}\right) \left[ d^L_{1m'}(\cos\theta)-(-1)^{m'}d^L_{1-m'}(\cos\theta)\right].
\eea
Eq.~\eqref{eq:Regge} is a special case of the general Regge symmetry~\cite{Regge:1958aa}, which allows exchange of the `lensing'  quantum number $M$ for $\Ml = \ell_2 - \ell_1$: if we define $\mathsf{m} = (m_1 - m_2)/2$, then
\begin{equation}\label{eq:Regge2}
\begin{pmatrix}
	L & \ell_1 & \ell_2 \\ -M & m_1 & m_2
\end{pmatrix} = \begin{pmatrix}
	L & \ell - \Ml/2 & \ell + \Ml/2 \\ -M & \mathsf{m} + M/2 & -\mathsf{m} + M/2
\end{pmatrix} = \begin{pmatrix}
	L & \ell - M/2 & \ell + M/2 \\ -\Ml & \mathsf{m} + \Ml /2 & -\mathsf{m} + \Ml/2
\end{pmatrix}.
\end{equation}
For the squeezed triangles with $L\ll \ell_1, \ell_2$ we have the following approximation consistent with the Regge symmetry
\begin{equation}
\label{eq:squeezed3j}
(-1)^{\ell_2-m_2}\threej{L}{\ell_1}{\ell_2}{M}{m_1}{m_2} \approx \frac{d^L_{M \Ml}(\theta)}{\sqrt{2\ell+1}}
\quad\text{or}\quad
\threej{L}{\ell_1}{\ell_2}{M}{m_1}{m_2} \approx (-1)^{\ell - \mathsf{m}} (-1)^{\frac{\Ml-M}{2}}\frac{d^L_{M \Ml}(\theta)}{\sqrt{2\ell+1}},
\end{equation}
where $\cos\theta \equiv 2\mathsf{m}/(2\ell+1)$. This also obeys the exact $(M,\Ml)\leftrightarrow (-M,-\Ml)$ symmetry of Eq.~\eqref{eq:Regge2} and hence cannot have odd $\mathcal{O}(\Ml/\ell)$ or $\mathcal{O}(M/\ell)$ corrections, giving higher-order accuracy that the original form of the approximation that does not respect the Regge symmetry~\cite{Brussaard57}.
\AL{I believe this holds to $O((L/\ell)^2)$ unlike approx seen elsewhere, this is as much as I can easily prove ..}

\subsection{Unnormalized polynomials}
We can also work with polynomials in both $\Ml$ and $L$ with integer coefficients, and leading coefficient following Chebyshev's, as follows:
\begin{equation}
		\mathcal P^L_m(\Ml) \equiv \sqrt{\frac{ (L + m)!}{(L - m)!} }\T^L_m(\Ml), \quad 	\mathcal Q^L_m(\Ml) \equiv \sqrt{\frac{ (L + m)!}{(L - m)!}} \frac{\U_m^L(\Ml)}{\sqrt{L(L + 1)}}.
\end{equation}
The recursion relation becomes
\begin{equation}
	\mathcal P^L_{m+1}(\Ml)=	2M \mathcal P^L_m(\Ml) - (L + m)(L - m + 1) \mathcal P^L_{m-1}(\Ml),
\end{equation}
the same for $\mathcal P$ and $\mathcal Q$.
The first few polynomials are
\begin{equation}
\begin{split}
	\mathcal P^L_0(\Ml) &= 1 \\
		\mathcal P^L_1(\Ml) &= \Ml \\
		\mathcal P^L_2(\Ml) &= 2 \Ml^2 - L(L+1)  \\ 
		\mathcal P^L_3(\Ml) &= 4 \Ml^3- \Ml (3 L^2 + 3 L - 2)\\
		\mathcal P^L_4(\Ml) &= 8 \Ml^4 -8\Ml^2 (L-1)(L + 2) + L(L + 1)(L-2)(L+3)\quad\\
		\vdots \\
\end{split}\:
\begin{split}
\mathcal Q^L_0(\Ml) &= 0 \\
	\mathcal Q^L_1(\Ml) &= 1 \\
		\mathcal Q^L_2(\Ml) &= 2 \Ml \\
		\mathcal Q^L_3(\Ml) &= 4 \Ml^2 - (L + 2)(L - 1)  \\
		\mathcal Q^L_4(\Ml) &= 8 \Ml^3 - 4\Ml (L^2 + L - 4)  \\
		\vdots \\
\end{split}
\end{equation}
in agreement with the Appendix of~\cite{Pearson:2012ba} for the parity-even case.
The general form for the parity-even case is (for $m>0$)
\be
\mathcal P^L_m(\Ml) = (-1)^{(L-\Ml)/2} \frac{\left((L+\Ml)/2\right)!\left((L-\Ml)/2\right)!}{(L-m)!} \sum_k  \left(-1\right)^{k} {\begm {L +m}\\{k}\enm} \begm {L -m}\\{\Ml -m +k}\enm.
\ee

\begin{figure}
    \centering
    \includegraphics[width=0.99\columnwidth]{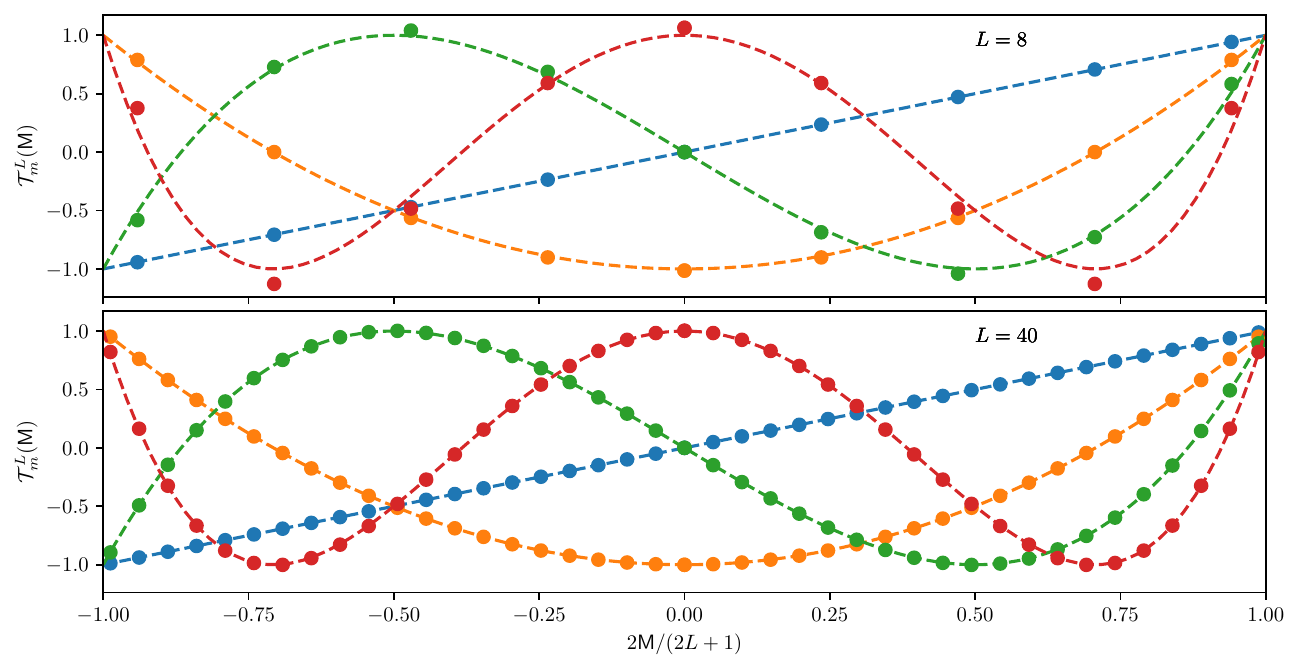}
    \caption{Comparison between the first few discrete polynomials $\mathcal T_m^L(\Ml =\ell_2 - \ell_1)$ introduced in this work, (points, for $m = 1, 2, 3, 4$), and the continuous Chebyshev polynomials $T_m(\cos \varphi) = \cos(m  \varphi)$ (dashed lines). The point abscissae correspond to $\cos \varphi = 2 (\ell_2 -
    \ell_1) / (2L + 1)$, with $\ell_2 -\ell_1$ running over the range allowed by the Wigner symbol, ($-L, L$ with $L + \ell_1 + \ell_2$ even). The upper panel shows $L=8$ and the lower panel $L = 40$.}
    \label{fig:Cheby}
\end{figure}

\section{GMV curved-sky quadratic estimators}\label{app:QEs}
In this section we review and discuss some aspects of the global minimum variance (GMV) curved-sky quadratic estimators. 

We start by justifying the result of Eq.~\eqref{QEweightbi}, that relates the quadratic estimator weights to the response bispectra. This also allow us to introduce some notation dealing with the minimal variance estimators. One quick way to obtain optimal quadratic estimator weights is through the gradient of the anisotropy source likelihood function. If the anisotropy is small, the likelihood is Gaussian, this will be optimal (the first step of a Newton-Raphson extremization, starting from vanishing anisotropy, goes right to the likelihood maximum), and we recover the usual quadratic estimator formalism. Thus, let
$\clmat_x^{XZ}\equiv \av{X_{\ell_1 m_1}Z^*_{\ell_2 m_2}}_{x}$ the observed two-point covariance for a fixed anisotropy source $x$. We can define the unnormalized estimator as
\begin{equation}
	\bar x_{LM} = - \frac{\delta}{\delta  x^*_{LM}}  \frac 12 \sum_{XZ, \ell_1 m_1,\ell_2 m_2} \: X^*_{\ell_1 m_1} \left[\clmat^{-1}_x \right]^{XZ}_{\ell_1 m_1, \ell_2 m_2} \, Z_{\ell_2 m_2} = \frac12 \sum_{XZ, \ell_1 m_1,\ell_2 m_2}\bar X^*_{\ell_1 m_1}\bar Z_{\ell_2 m_2} \frac{\delta}{\delta x^*_{LM}}\av{X_{\ell_1 m_1}Z^*_{\ell_2 m_2}}_{x}
\end{equation}
where we defined the inverse-variance filtered harmonic modes $\bar X = \sum_Z \left[\clmat_{x=0}^{-1}\right]^{XZ} Z$. The definition of the $x_{LM}$ response function consistent with Eq.~\eqref{lenskernel} for fixed $m_1$ and $m_2$ is
\begin{equation}
	\av{X_{\ell_1 m_1} Z_{\ell_2 m_2}}_{x} \ni \sum_{LM}\begin{pmatrix}
		L & \ell_1 & \ell_2 \\ M & m_1 & m_2
	\end{pmatrix} f^{x XZ}_{L \ell_1 \ell_2} x^*_{LM}.
\end{equation}
Complex conjugating $X_{\ell_1 m_1} $ and $\bar X_{\ell_1 m_1}$, and inserting the complex conjugated response into the estimator, we get
\begin{align}\label{eq:appGMV}
	\bar x_{LM} &= \frac {(-1)^M}2 \sum_{XZ,\ell_1 m_1, \ell_2 m_2} \begin{pmatrix}
		L & \ell_1 & \ell_2 \\ -M & m_1 &  m_2  
	\end{pmatrix}\bar X_{\ell_1 m_1}\bar Z_{\ell_2 m_2}  \left[f^{x X Z}_{L \ell_1 \ell_2} \right]^*.
\end{align}
Comparing with the definition of Eq.~\eqref{eq:QEdef}, we can read off the optimal weights, $f^*$. Including non-perturbative terms from Eq.~\eqref{lenskernel} (i.e. the effect of other lensing modes at different $L'M'$) improves the estimator~\cite{Lewis:2011fk}.

We now discuss some further properties of the GMV lensing gradient and curl estimator, starting from its position-space formulation. Our goals in doing so are i) obtain explicit expression for the weights appearing in Eq.~\eqref{eq:appGMV}, ii) motivate our bispectrum reduction procedure in the case of the parity-odd combinations, iii) discuss briefly practicalities affecting the implementation of our results.

The lensing estimators take a very simple form in position space, but there are an array of slightly different notations that can make comparison confusing. We adopt here notation that respects standard conventions for spin-weighted spherical harmonic transforms:  the gradient and curl modes $G$ and $C$ of a spin-s field $_{s}f$ are defined through
\begin{equation}
	{}_{\pm s} f(\hn) =   \sum_{\ell m} \:_{\pm s}{f}_{\ell m} \: _{\pm s}Y_{\ell m}(\hn)
\equiv - (\pm)^{s} \sum_{\ell m} (G_{\ell m} \pm i C_{\ell m}) \: _{\pm s}Y_{\ell m}(\hn). \quad (s \geq 0),
\end{equation}
and the spin-raising and lowering operators are
\begin{equation}
	\eth^\pm \:_sf = -(\sin \theta)^{\pm s} \left[ \frac{\partial }{\partial \theta} \pm \frac{i}{ \sin \theta} \frac{\partial}{\partial \phi} \right]\left[ (\sin \theta)^{\mp s} \:_sf \right].
\end{equation}
These definitions gives standard polarization $E$ and $B$ modes from spin-2 polarization $_{\pm 2} P(\hn) = Q(\hn) \pm i U(\hn)$, but the spin-0 gradient mode sign can be counter-intuitive: the gradient mode of $T(\hn)$ or of the lensing potentials $\psi(\hn)$ and $\Omega(\hn)$ are respectively $-T_{\ell m}$, $-\psi_{LM}, -\Omega_{L M}$. This distinction for $T$ is inconsequential to this appendix, since $T$ always enter either squared or with a factor of $\tilde C_\ell^{XT}$ for other $X$'s. A pure gradient field of the form $\eth^+{}_{0}f$ has gradient mode $\sqrt{\ell (\ell + 1)} G_{\ell m}$. Owing to the minus sign in the definitions of the spin-raising operator, the pure gradient lensing deflection vector field $_1d(\hn) = d_\theta(\hn) + i d_\phi(\hn)$ is instead $-\eth^+\psi(\hn)$, so that there is the expected correspondence $d_\theta = \partial_\theta \psi$ and $d_\phi =   \partial_\phi \psi / \sin \theta $,  giving the standard gradient on the flat-sky with coordinates $\theta$ and $\phi$. The gradient and curl modes of the deflection field are $\sqrt{L (L + 1)} \psi_{LM}$,  $\sqrt{L (L + 1)} \Omega_{LM}$. 

Following Ref.~\cite{Carron:2017mqf, Maniyar:2021msb, Aghanim:2018oex}, an un-normalized optimal quadratic estimator for the lensing deflection vector field that is built from a set of fields with spins labeled with $s$ (typically $s \in (0, \pm 2)$, but results in this section holds for any $s$) can be written as
\newcommand{\thickbar}[1]{\accentset{\rule{.5em}{1pt}}{#1}}
\begin{equation}\label{eq:d1}
	_{\pm1} \hat d(\hn)  = - \sum_{s}\:_{-s} \thickbar{X} (\hn)\eth^{\pm}{} _{s}X^{\rm WF}(\hn) = - (\pm)  \sum_{LM} \left (\frac{\bar \psi_{LM} \pm i \bar \Omega_{LM}}{\sqrt{L(L + 1)}} \right) \:_{\pm 1 }Y_{LM}(\hn),
\end{equation}
where $X^{\rm WF}$ is the Wiener-filtered CMB. The maps in this equation are obtained from the inverse-variance filtered harmonics through the relations
\begin{equation}\label{eq:wfres}
X_{\ell m}^{\rm WF} = \sum_Z\tilde{\boldsymbol{C}}_{\ell}^{XZ} \bar Z_{\ell m},\quad {}_{\pm s} \thickbar X(\hn) \equiv \frac{\:_{\pm s}\bar X(\hn)}{2 - \delta_{s0}} = - (\pm)^{s}  \sum_{lm} \left(\frac{\bar G_{\ell m} \pm i \bar C_{\ell m}}{2 - \delta_{s0}} \right) \: _{\pm s}Y_{\ell m}(\hn)
\end{equation}
Most references including \emph{Planck}~\cite{Aghanim:2018oex} and ACT DR6~\cite{ACT:2023dou} lensing, as well as Refs.~\cite{Maniyar:2021msb, Belkner:2023duz} use the notation ${}_{s}\bar X(\hn)$ on \eqref{eq:d1}, instead of our ${}_{s}\thickbar X(\hn)$, where our thick overbar explicitly denotes filtering the complex field in real space, rather than filtering the component modes in harmonic space. There is a relative factor of 2 in Eq.~\eqref{eq:wfres} for non-scalar complex transforms between real and harmonic space (see for example Section 5.7 of~\cite{ACT:2023dou} for the origin of this factor of 2; for idealized uncorrelated noise it is simply that $\la {}_{\pm2}P {}_{\pm 2}P^*\ra = 2\la Q^2\ra= 2\la U^2\ra$, so inverse-filtering with the complex variance gives half the result of inverse-filtering the real $Q$ and $U$ with their variances).

For best QE results, and on an otherwise isotropic configuration, the matrix $\tilde{\boldsymbol{C}}_{\ell} $ is made out of the non-perturbative gradient response spectra, but we used the lensed spectra for numerical results in this paper for simplicity. From Eq.~\eqref{eq:d1},  we can equivalently write
\begin{equation}\label{eq:gpgO}
	\bar \phi(\hn)  = \frac{1}{2}\left(\eth^- \:_{+1} d(\hn) +  \eth^+ \:_{-1} d(\hn) \right), \quad \bar \Omega(\hn)  = \frac{1}{2i}\left(\eth^- \:_{+1} d(\hn) - \eth^+ \:_{-1} d(\hn) \right),
\end{equation}
which we can then obtain in harmonic space with a spin-0 transform.

We can proceed now to write the estimator in the form of Eq.~\eqref{eq:appGMV}. To do this, we apply the chain rule to Eq.~\eqref{eq:gpgO} and go to harmonic space with a spin-0 transform. 
With the understanding that $X^s_{\ell m}$, $Z^{s,\rm WF}_{\ell m}$ stands for either the gradient or curl mode of the corresponding spin-s map, the estimator results in
\newcommand{\ired}[0]{\color{red}i\color{black}\xspace}
\newcommand{\cyan}[1]{\color{cyan}#1\color{black}\xspace}
\newcommand{\red}[1]{\color{red}#1\color{black}\xspace}
\newcommand{\olive}[1]{\color{olive}#1\color{black}\xspace}
\newcommand{\blue}[1]{\color{blue}#1\color{black}\xspace}

\begin{equation}
\begin{split} \label{eq:Xs}
	\bar x_{LM}& =  \frac 1{2\ired}\sum_{\ell_1 m_1, \ell_2m_2}\sum_{s \ge 0, XZ}\bar X^s_{\ell_1 m_1} Z^{s, \rm WF}_{\ell_2 m_2}	 (-1)^{M} \begin{pmatrix}
		\ell_1 &  \ell_2 & L \\ m_1 & m_2 &  -M
	\end{pmatrix} \sqrt{ \frac{(2\ell_1 + 1)(2\ell_2 + 1)(2L + 1)}{4\pi}} 
	\\ & \times \: (-1)^s \left[ \;_{s+1}F^{\pm}\ell_1^{(s+1)}\ell_2^{(s+1)}  + \;\red{(-1)} _{s-1}F^{\pm}\ell_1^{(s)}\ell_2^{(s)} +\:_sF^{\pm}\left( \ell_2^{(s+ 1)} \ell_2^{(s + 1)} +\red{(-1)}\ell_2^{(s)} \ell^{(s)}_2  \right) \right]\cyan{(-i)}\blue{(i)},
\end{split} 
\end{equation}
where \red{red} factors apply to \red{curl} mode reconstruction only, \cyan{cyan} factors to the \cyan{$\bar C_{\ell_1 m_1} G_{\ell_2 m_2}^{\rm WF}$ } part only, \blue{blue} factors to the \blue{$\bar G_{\ell_1 m_1} C_{\ell_2 m_2}^{\rm WF}$ } part only, and the $\pm$ on the $F$'s are for even and odd field combinations respectively. We defined
\begin{equation}
 _sF^{\pm} \equiv \frac12 \begin{pmatrix}
		L & \ell_1 & \ell_2 \\ 0 & -s &  s 
	\end{pmatrix}  \pm \: \frac 12 \begin{pmatrix}
		L & \ell_1 & \ell_2 \\ 0 & s &  -s 
	\end{pmatrix} \text{   as well as   } \ell^{(s + 1)} \equiv \sqrt{\ell(\ell+ 1) - s(s+1)} = \sqrt{(\ell- s)(\ell + s + 1)},
\end{equation}
the response of the spin-s spherical harmonic to the spin-raising operator ($\eth^+ {}_s Y_{\ell m}=\ell^{(s+ 1)}{}_{s + 1}Y_{\ell m}$, with $\ell^{(1)} = \ell^{(0)}  = \sqrt{\ell(\ell + 1)}$). The response of the spin-s harmonic to the spin lowering estimator is negative, which brings an overall minus sign to that equation ($\eth^- {}_s Y_{\ell m}=-\ell^{(s)}{}_{s - 1}Y_{\ell m}$).
Alternative ways of writing the terms in this general result can be found in Ref.~\cite{Namikawa:2011cs}.

The $F$ symbols obey a simple three-term recursion relation inherited from the constituent Wigner 3j:
\begin{align}
&	_s{F}^{\pm} \left[L(L + 1) - \ell_1 (\ell_1 + 1) - \ell_2 (\ell_2 + 1) + 2s^2  \right] =  \:_{s-1}{F}^{\pm}{\ell_1}^{(s)}{\ell_2}^{(s)} + \:_{s+1}F^{\pm}{\ell_1}^{(s+1)}{\ell_2}^{(s+1)}.
\end{align}
This recursion is initiated with the first two terms
\begin{equation}\label{eq:apprec}
_0F^+ = \begin{pmatrix}
	L & \ell_1 & \ell_2 \\ 0 & 0 & 0 
	\end{pmatrix},	
\quad\: _0F^- =0, \quad	(-1)\:_{\pm 1} F^+  = \textbf{cos}\:\theta_L  \: \begin{pmatrix}
	L & \ell_1 & \ell_2 \\ 0 & 0 & 0 
	\end{pmatrix},\quad 	_{\pm 1} F^-  = \pm\,\textbf{sin}\:\theta_L  \: \begin{pmatrix}
	L-1 & \ell_1 & \ell_2 \\ 0 & 0 & 0 
	\end{pmatrix}
\end{equation}
where
\begin{align}\label{eq:apprec2}
&\textbf{cos} \:\theta_{L} \equiv  \frac{\ell_1(\ell_1 + 1)  +\ell_2(\ell_2 + 1) - L(L + 1) }{2 \sqrt{\ell_1(\ell_1 + 1)\ell_2(\ell_2 + 1)}} \quad \textbf{sin}(\theta_{L}) \equiv \frac{ \sqrt{L^2 - (\ell_1-\ell_2)^2} \sqrt{ (\ell_1 + \ell_2 + 1)^2 - L^2}}{2 \sqrt{\ell_1(\ell_1 + 1)\ell_2(\ell_2 + 1)}}
\end{align}
are quantized analogs of the cosine and sine of $\theta_L$ in the triangle defined by the three momenta (the result for $_{\pm 1}F^+$ follows directly from the recursion relation, and for $_{\pm 1} F^-$ from the explicit form for the Wigner functions~\cite{Pain:2020tid}). More generally, the ratios $(-1)^s \:_sF^{+}/_0F^{+}$ and $(-1)^{s-1} \:_sF^{-} /\: _1F^{-}$ are quantized versions of the trigonometric factors $\cos (s\theta_L)$ and $\sin (s\theta_L)/\sin \theta_L$, to which they reduce for large momenta: this may be seen from the recursion relation, which becomes that of the Chebyshev polynomials. 
In that regime, up to $(L/\ell)^2$ and $(s/\ell)^2$ corrections (but without other constraints on $L$), we may write
\begin{align}
\label{eq:squeezedF}
(-1)^s{}_{s}F^{+} \approx {}_{0}F^+ \approx (-1)^{\frac{L + 2\ell}{2}}\frac{|\Delta^L_{0\Ml}|}{\sqrt{2\ell + 1}}\quad\text{  and  }\quad
(-1)^{s-1}\:{}_{s}F^{-} \approx s \: {}_{1}F^- \approx (-1)^{\frac{L + 2\ell + 1}{2}}\frac{2s}{2\ell + 1} \sqrt{L(L + 1)} \frac{|\Delta^L_{1\Ml}|}{\sqrt{2\ell + 1}},
\end{align}
consistent with Eq.~\eqref{eq:squeezed3j}.

On the practical side, it is clear from the recursion that $_sF^+ / _0F^+$ is a polynomial in $L(L + 1)$ of order $s$, and  $_sF^- / _1F^-$ a polynomial in $L(L + 1)$ of order $s-1$. We have used this property to generate many of the numerical results of this work -- since the reduced weight function becomes a polynomial, numerical evaluation of multipole coefficients can proceed efficiently order by order where the $L$-dependency can be factorized out. In the gradient case, the $F$ terms in \eqref{eq:Xs} always combine using the recursion formulae for the $F$'s, so the term in the square bracket of \eqref{eq:Xs} simplifies for any $s$ to
\begin{equation}
	_{s}F^{\pm}\left(L(L+1) -\ell_1(\ell_1 + 1) + \ell_2 (\ell_2 + 1) \right) \quad \quad\text{   (square bracket of \eqref{eq:Xs}, gradient case)}
\end{equation}
This corresponds to integrating by parts the relevant terms in well-known derivations of the weights~\cite{Okamoto:2003zw, Challinor:2002cd, Namikawa:2011cs})\JC{Would be nice to simplify the curl as well}
We can see from Eq.~\eqref{eq:apprec} and \eqref{eq:apprec2} that $\sqrt{L^2 - \Ml^2}\begin{pmatrix}
	L-1 & \ell_1 & \ell_2 \\ 0 & 0 & 0 
	\end{pmatrix}$ appears in a fairly natural way in all parity-odd lensing estimators, as claimed in the main text. Dividing by this factor leaves behind a much simpler $L$-dependence, in analogy to the even-parity bispectrum reduction. More generally, this happens for all separable parity-odd bispectra with Gaussian $L$-mode~\cite{Namikawa:2011cs, Kamionkowski:2010rb}.

\section{Large-lens limits}\label{app:sqz}
This appendix tabulates the curved-sky low-$L$ behaviour of the gradient and curl lensing estimators, correct to $( L / \ell	)^2$. 
These results can be obtained from the kernels in appendix~\ref{app:QEs}, by series expanding and computing the corresponding moments. 
This is, however, slightly tedious work. Hence, in the second part of this appendix we also give a simpler flat-sky based argument that motivates and justifies all the prefactors.

For simplicity, we give here the result for the case where the filtering of temperature and polarization is made independently (neglecting $C_\ell^{TE}$ in $\boldsymbol{C}^{-1}$, resulting in a diagonal matrix; this has been called the SQE estimator by~\cite{Maniyar:2021msb}). In the most general case, the result must be adapted slightly, as described further below. 
\begin{align} \nonumber 
	\mathcal R_L^{\kappa, XY} &\sim \frac {2 - \delta_{XY}}2 \sum_\ell \left ( \frac{2\ell + 1} {4\pi} \right) \left(\frac{\Cgrad^{XY}_\ell}{C_\ell^{XX,\rm tot}}\frac{\Cgrad^{XY}_\ell}{C_\ell^{YY,\rm tot}} \right)\left[\left(\frac{\d \ln \ell(\ell+1)\Cgrad^{XY}_\ell}{\d \ln (2\ell+1)} \right)^2  + \frac 12 \frac{(L-1)(L + 2)}{L(L + 1)}\left(\frac{\d \ln \Cgrad^{XY}_\ell}{\d \ln (2\ell+1)} \right)^2\right] \\
		\mathcal R_L^{\omega, XY} &\sim \frac {2 - \delta_{XY}}2 \sum_\ell \left ( \frac{2\ell + 1} {4\pi} \right) \left(\frac{\Cgrad^{XY}_\ell}{C_\ell^{XX,\rm tot}}\frac{\Cgrad^{XY}_\ell}{C_\ell^{YY,\rm tot}} \right)\left[ \frac 12 \frac{(L-1)(L + 2)}{L(L + 1)}\left(\frac{\d \ln \Cgrad^{XY}_\ell}{\d \ln (2\ell+1)} \right)^2\right] \label{eq:sqzall} \\ 
		\text{for } XY &\in (TT, TE, EE, BB), \text{and} \nonumber \\
					\mathcal R_L^{\kappa, EB}&\sim   \sum_\ell \left ( \frac{2\ell + 1} {4\pi} \right) \left(-2\frac{\tilde C^{EE}_\ell - \tilde C^{BB}_\ell}{C_\ell^{EE,\rm tot}} \right)\left(-2\frac{\tilde C^{EE}_\ell - \tilde C^{BB}_\ell}{C_\ell^{BB,\rm tot}} \right)\left[\frac 12 \frac{(L-1)(L + 2)}{L(L + 1)} \right] \nonumber\\
					\mathcal R_L^{\omega, EB}&\sim   \sum_\ell \left ( \frac{2\ell + 1} {4\pi} \right) \left(-2\frac{\tilde C^{EE}_\ell - \tilde C^{BB}_\ell}{C_\ell^{EE,\rm tot}} \right)\left(-2\frac{\tilde C^{EE}_\ell - \tilde C^{BB}_\ell}{C_\ell^{BB,\rm tot}} \right)\left[1 + \frac 12 \frac{(L-1)(L + 2)}{L(L + 1)} \right] \nonumber \\
						\mathcal R_L^{\kappa, TB}&\sim   \sum_\ell \left ( \frac{2\ell + 1} {4\pi} \right) \left(-2\frac{\tilde C^{TE}_\ell}{C_\ell^{TT,\rm tot}} \right)\left(-2\frac{\tilde C^{TE}_\ell }{C_\ell^{BB,\rm tot}} \right)\left[\frac 12 \frac{(L-1)(L + 2)}{L(L + 1)} \right] \nonumber\\
					\mathcal R_L^{\omega, TB}&\sim   \sum_\ell \left ( \frac{2\ell + 1} {4\pi} \right) \left(-2\frac{\tilde C^{TE}_\ell }{C_\ell^{TT,\rm tot}} \right)\left(-2\frac{\tilde C^{TE}_\ell }{C_\ell^{BB,\rm tot}} \right)\left[1 + \frac 12 \frac{(L-1)(L + 2)}{L(L + 1)} \right]. \nonumber
\end{align}
For $TE$, since the spectrum can oscillate through zero, log-derivatives like $C_\ell \:d \ln C_\ell $ should be understood as $dC_\ell$.
In the first line, the first term in the square brackets corresponds to the monopole ($m=0$, magnification), and the second term, with the slight $L$-dependence at low $L$, to the quadrupole ($m=2$, shear). Other multipoles are suppressed by $(L / \ell)^2$.
The lensing curl mode response for the first set of combinations is shear-only, and the same as the shear-only contribution of the lensing gradient mode, and vice-versa for the second set of combinations, in which case the lensing gradient mode is shear-only.

A simple way to understand the prefactors and derive these formulae is with the following argument: low $L$'s vary slowly in real space, changing the local, 2-dimensional power spectra $C_{\vl}$. In the regime of small signals, the quadratic estimators by construction capture all the information available. For this reason,  in the low $L$ limit, the response of a quadratic estimator with optimal weights targeting anisotropy source $\alpha$ to anisotropy parameter $\beta$ will be given by the Fisher information matrix formula
\begin{equation}\label{eq:Fisher}
	\mathcal R_L^{\alpha \beta} \approx \frac 12 \sum_{\vl} \textrm{Tr}\left[\frac{\partial C_{\vl}}{\partial \alpha } \boldsymbol{C}^{-1}_l\frac{\partial  C_{\vl}}{\partial \beta} \boldsymbol{C}^{-1}_l \right].
\end{equation}
Here $\alpha$ and $\beta$ should be understood as the local parameters of interests as sourced by the global field. For lensing, these are the four elements of the magnification matrix~\cite{Lewis:2006fu}\footnote{Our full-sky definitions give a different sign to $\omega$ in the flat-limit, compared to that reference. In our conventions, an image appears to the observer rotated \emph{anti-clockwise} by the angle $\omega$.} that describes the local remapping $\boldsymbol{\theta} \rightarrow \boldsymbol{\theta} + \boldsymbol{d}(\theta)$
\begin{equation}
	A_{ij} - \delta_{ij}  = \frac{\partial d_j}{\partial \theta_i} = \begin{pmatrix} 
	- \kappa - \gamma_1 & -\gamma_2 - \omega \\ - \gamma_2 + \omega & - \kappa + \gamma_1
	\end{pmatrix}.
\end{equation}
The sum in \eqref{eq:Fisher} runs over the modes of the local patch. The spectra responses are anisotropic for non-scalar sources like the shears.
The response of the spectra $\partial_\alpha C(\vl)$ (listed below for completeness) can be obtained in the flat-sky approximation, for example from the change in the local correlation function under the induced change of separation $\vr \rightarrow (1 + A)\vr $. 

Hence, there is a 3 steps strategy to obtain the full-sky low-$L$ response of a quadratic estimator: i) compute these spectra derivatives ii) remap $\alpha$ and $\beta$ to the harmonic modes of the global fields of interest (this involves computing things like $\partial \gamma / \partial\kappa_{LM}$, and similarly for  the local convergence and rotation) ii) sum up the information in all patches.

With this in place, the various prefactors comes about in the following way: the shear prefactor $(L-1)(L + 2)/L(L + 1)$ is the proportionality constant between the squared shear field full-sky harmonic modes $\gamma_{LM}$ and that of the convergence $\kappa_{LM}$ (step ii) as described above): the full-sky relations
\begin{equation}
	\gamma_1(\hn) + i \gamma_2(\hn) = -\frac 12 \eth^+\eth^+ \: (\phi(\hn) +i \Omega(\hn)), \quad	\kappa(\hn) + i \omega(\hn) = -\frac 12  \eth^-\eth^+ \:(\phi(\hn) +i\Omega(\hn))
\end{equation}
imply
\begin{align}
	_2|\gamma^{\text{E-mode}}|_{LM}^2 &= \frac{(L-1)(L + 2)}{L(L + 1)} |\kappa_{LM}|^2, \quad  _2|\gamma^\text{B-mode}|_{LM}^2= \frac{(L-1)(L + 2)}{L(L + 1)} |\omega_{LM}|^2.
\end{align}
There is an additional $1/2$ factor in front of this factor in \eqref{eq:sqzall}, because there are two independent local shear components (and hence the response, or inverse variance, decreases by two). The $(4\pi)^{-1}$ appears converting the position space $\kappa$ and $\omega$ to the full-sky harmonic modes (also step ii). Finally, the factor $2-\delta_{XY}$ comes about because there are two ways to produce $\tilde C_\ell^{XY}$ in the trace~\eqref{eq:Fisher} under the stated assumption of diagonal filtering, but only one way for $X = Y$. In the $TB$ and $EB$ terms, we have left factors of $-2$ within the response spectra, which is where they originate. They combine to the prefactor given in the main text~\eqref{eq:EBsqdz}. The GMV limits instead of SQE may be obtained dropping the assumption of diagonal filtering in \eqref{eq:Fisher}.

Finally, for completeness, writing $\vl = le^{i \phi_{\vl}}$, the flat-sky responses are
\begin{align}\nonumber
\delta C_{\vl}^{XX} &\sim  \kappa \frac{d \ln l^2 C^{XX}_l}{d\ln l} + (\gamma_1 \cos(2\phi_{\vl}) + \gamma_2 \sin(2\phi_{\vl}))\frac{d\ln C^{XX}_l}{d\ln l}	\quad \text{ for $XY \in (TT, EE, TE, BB)$} 
\\  \delta C_l^{EB} &\sim  -\left(C_l^{EE} - C_l^{BB}\right)2\left[\omega + \gamma_1 \sin(2\phi_{\vl}) - \gamma_2 \cos(2\phi_{\vl}) \right],\quad \delta C_l^{TB} \sim -C_l^{TE} \:2 \left[ \omega  + \gamma_1 \sin(2\phi_{\vl}) - \gamma_2 \cos(2\phi_{\vl})\right]
\end{align}
A useful reference for flat-sky lensing kernels is~\cite{Sailer:2022jwt}.
With this argument one can also easily obtain the squeezed limits for estimators targeting other sources of anisotropies.

\section{ Evaluation of the non-separable quadratic estimators}

We now briefly discuss our strategy for evaluating the moment estimators. Unlike the standard estimators, they are not easily separable, and hence have additional computational cost. But it is still doable, as described here for one possible method that is suitable for moderate values of $m$, which are the most interesting. For simplicity, we describe here the case of parity even estimators, like $TT$, $EE$, etc, which can all be processed in the same way. The treatment of parity-odd estimators only requires minor adjustements, by going from spin 0 to spin 1 transforms in what follows.

We must evaluate estimators of the type given in Eq.~\eqref{eq:QEm}, which is for each $L$ and $M$ a non-separable sum over 3 indices. In the case of the original estimators, the separability in $\ell_1$ and $\ell_2$ allows fast evaluation in the position-space, after writing the 3j symbols in their Gaunt integral formulation. This suggests reinserting an explicit full Gaunt factor:
\begin{equation}
	\bar x^{(m)}_{LM} =\frac{(-1)^M}{2} (2 - \delta_{0m})\!  \sum_{\ell_1 m_1, \ell_2 m_2}  \mathcal T^L_{m}(\Ml)\:\tilde w^{(m)}_{L\ell}\tilde X_{\ell_1 m_1}\tilde X_{\ell_2 m_2}\: \begin{pmatrix}
		L & \ell_1 & \ell_2 \\ -M & m_1 & m_2
	\end{pmatrix}  \begin{pmatrix}
		L & \ell_1 & \ell_2 \\ 0 & 0 & 0
	\end{pmatrix}\sqrt{ \frac{(2\ell_1 + 1)(2\ell_2 + 1)(2L + 1)}{4\pi}} .		
\end{equation}
In this equation, we defined
\begin{equation}
	\tilde X_{\ell m} \equiv \bar X_{\ell m} \sqrt{\frac{(4\pi)^{1/2}}{(2\ell + 1)\bar C_\ell}}, \quad \tilde w^{(m)}_{L\ell} \equiv \frac{w_{L\ell}^{(m)}|\Delta_{0\Ml}|}{\sqrt{2L + 1}\begin{pmatrix}
		L &  \ell_1 & \ell_2 \\ 0 & 0 & 0 
	\end{pmatrix}}.
\end{equation}
We can now restore separability with the following two Fourier `transforms',
\begin{equation}
		\mathcal T^L_m(\Ml) =\left. \mathcal{T}_m^L\left(i \frac{d}{d\psi}\right)\right|_{\psi = 0} e^{-i (\ell_2 - \ell_1)\psi },\text{  and  } \tilde	w^{(m)}_{L \ell} \equiv \int_0^{2\pi} \frac{d\phi }{\sqrt{{2\pi}}} e^{- i (\ell_1 + \ell_2) \phi}\: \tilde w^{(m)}_L(\phi).
\end{equation} 
The first one works because $\mathcal T^L_m$ is a polynomial in $\Ml$, and the second one is in practice a Fast Fourier Transform with $\ell_{\rm max}$ points. Carrying out these steps eventually gives 
\begin{equation}
		\bar x^{(m)}_{LM} =\frac 12(2 - \delta_{m0})\: \left. \mathcal{T}_m^L\left(i \frac{d}{d\psi}\right)\right|_{\psi = 0} \int_0^{2\pi} \frac{d\phi}{\sqrt{2\pi}}\: \tilde w^{(m)}_L(\phi) \left[\tilde T(\phi - \psi )\tilde T(\phi +\psi) \right]_{LM}.
		\end{equation}
In this equation $\tilde T(\phi \pm \psi)$ is a position space map with harmonic coefficients $\tilde T_{\ell m}e^{-i \ell (\phi \pm \psi)}$; explicitly $\tilde T (\phi \pm \psi, \hn) = \sum_{\ell m} \tilde T_{\ell m}e^{-i \ell (\phi \pm \psi)} Y_{\ell m}(\hn)$.
 The first two moments in this representation are given by
\begin{align}\label{eq:TTnosep}
	\bar x^{(0)}_{LM}& = \frac 12 \int_0^{2\pi}\frac{ d\phi}{\sqrt{2\pi}} \:\tilde w^{(0)}_L(\phi) \left[ \tilde T^2(\phi) \right]_{LM}\\
	\bar x^{(2)}_{LM} &=  \sqrt{\frac{(L-2)!}{(L + 2)!}} \int_0^{2\pi} \frac{d\phi}{\sqrt{2\pi}}\tilde w^{(2)}_L(\phi) \left[   4\tilde T^{',2} (\phi) - 4 \tilde T^{''}(\phi)\tilde T(\phi) - L(L + 1) \tilde T^2(\phi)\right]_{LM}
\end{align}
where $'$ is derivative with respect to $\phi$.

This method just accumulates the spherical harmonic coefficients of a product of two maps over many $\phi$'s. For a given $\phi$, this has the same cost of a standard temperature-based quadratic estimator starting from the harmonic coefficients of two maps. There is optimized software now that can compute this in less than one second on the 4 laptop CPUs with $\ell_{1,2,\rm max}= 3000$, a reasonable maximal multipole for Simons Observatory for example. Since the parity of $2\ell = \ell_1 + \ell_2$ is that of $L$, it holds that $\tilde w^m_L(\pi +\phi) = (-1)^L \tilde w^m_L(\phi)$ as well as $\tilde w_L^m(\pi - \phi) = (-1)^L\left(\tilde w^m_L(\phi)\right)^*$, so that only a fourth of the $\phi$ points are relevant. The number of $\phi$ points needed is thus about $\ell_{\rm max} / 2$. We conclude that the total cost (for a single $m$) of evaluating the monopole fully is about 1 hour on a laptop, and obviously much less with the help of HPC resources, since the $\phi$ integral is embarrassingly parallel. In practice, many fewer points are needed to get the estimate to reasonable accuracy. Contributions are concentrated close to $\phi = 0$, with another bump at a scale corresponding to the position-space acoustic scale. For a SO-like configuration, this corresponds on the FFT grid to the index $i_{\phi}\sim 10$, so that the effective number of points needed is much smaller. See the right panel of Fig.~\ref{fig:bandlimit}.

\section{Impact of band limits} 
\label{bandlimits}
\newcommand{\ellmax}[0]{\ensuremath{\ell_{\rm max}}\xspace}

\begin{figure}[h!]
    \centering
    \includegraphics[width=0.42\columnwidth]{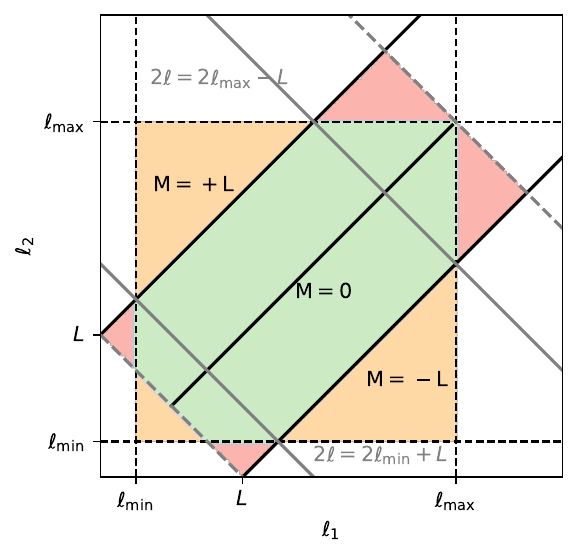}
        \includegraphics[width=0.53\columnwidth]{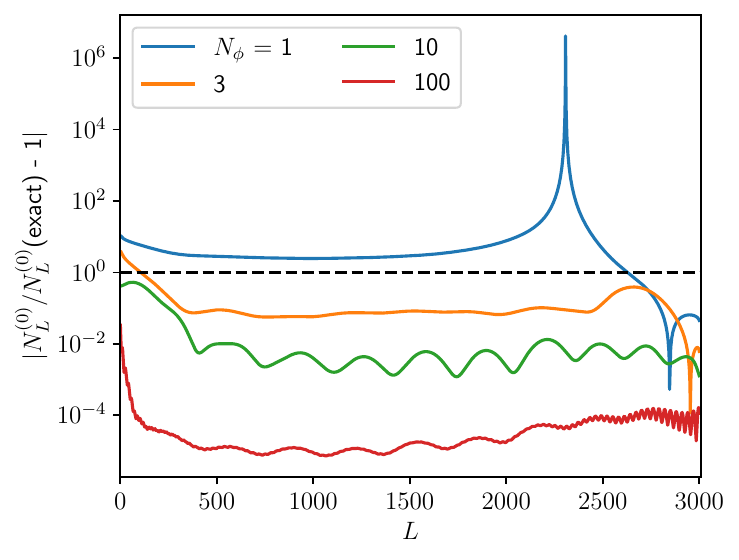} 

    \caption{\emph{Left panel:} Illustration of the effect of band limits on orthogonality of the quadratic estimators/bispectra. Ascending diagonals are lines of fixed $\Ml$, and falling diagonals lines of fixed $2\ell$. For a fixed $L$, the relevant set of $\ell_1$ and $\ell_2$ modes form an ascending band of width $2L + 1$ starting at $2\ell = L$. Fixing $2\ell$, not all $\Ml'$s with $|\Ml| \le L$ are always available in the presence of cuts $\ell_{\rm min}$ and $\ell_{\rm max}$ (the red areas fall outside of the cuts), making the bispectra not perfectly orthogonal. \emph{Right panel:} Comparison of reconstruction noise levels of the TT monopole estimator, as a function of the number of $\phi$-points used to compute the estimator (see Eq.~\eqref{eq:TTnosep}), starting from $\phi=0$. The fast convergence means that it is simple to produce this estimator even though it is not separable. Tests on actual maps confirm these findings.}
    \label{fig:bandlimit}
\end{figure}
In practice, there is a finite band limit, \ellmax, to the CMB maps used to construct quadratic estimators or bispectrum estimators. For a fixed lens multipole $L$ and CMB mode $2\ell = \ell_1 + \ell_2$, such that $2\ell \ge 2\ellmax -L$,  the band limit imposes a cut on $\Ml = \ell_2 - \ell_1$, such that $|\Ml| \le 2\ellmax - 2\ell$. See Fig.~\ref{fig:bandlimit}. Hence, for modes such that $L > 2\ellmax - 2\ell$, the overlap sum of a pair of polynomials is incomplete, and they are not exactly orthogonal anymore. This is a small correction, unless $L$ itself is comparable to the band limit so that significant information remains at higher $\ell$. The correction is easy to include by default in practice, and our implementation always includes these effects. A qualitatively similar, but quantitatively smaller correction also applies in the presence of a low-$\ell$ cut-off, with $|\Ml|$ bounded by $2\ell - 2\ell_{\rm min}$.

Allowing for a general rescaling by $h_{\ell_1}h_{\ell_2}$ of the weights that are expanded  (used for example in our discussion of the point-source deprojection), we compute the multipole coefficients as discussed in the main text (here for the parity even case)
\begin{equation}\label{eq:wdef}
	w_{L\ell}^{(m),h} = \avM{\frac{h_{\ell_2}h_{\ell_2}W_{L \ell_1 \ell_2}}{|\Delta^L_{0\Ml}|}\mathcal T^L_m(\Ml)}.
\end{equation}
In principle, one could extrapolate the relevant functions outside of the band limits to produce the weights independently of the band limits. We choose instead to use the band-limited functions in this equation. Hence the band limits affect the weights.

The unnormalized reconstruction noise entering in Equations~\eqref{eq:n0} or \eqref{eq:n0GMV} becomes
\begin{align}\label{eq:respgrad}
	n_L^{(0) mm'} = \frac {(2 -\delta_{0m}) (2 -\delta_{0m'})}{2(2L + 1)}\sum_{\ell}w^{(m),h}_{L\ell} w^{*,(m'),h}_{L\ell} \avM{\frac{\bar C_{\ell_1}}{|h_{\ell_1}|^2}\frac{\bar C_{\ell_2}}{|h_{\ell_2}|^2} \mathcal T^L_m(\Ml)\mathcal T^L_{m'}(\Ml)}.
\end{align}
To compute normalized reconstruction noises $N^{(0)}_L$, we also need the responses of the multipole estimators to the lensing (or other) signal. The response function is the complex conjugate of the weights $W_{L\ell_1\ell_2}$ \eqref{QEweightbi}. Hence, when computing the response, the polynomial $\mathcal T_m^L$ of the multipole estimator combines with the response function to produce $w_{L\ell}^{*(m),1/h}$. The form of the result is unaffected by the band-limits,\begin{equation}
 \mathcal R_L^{\kappa(m)} = \frac {(2 -\delta_{0m}) }{2(2L + 1)}\sum_{\ell}w_{L\ell}^{(m), h} w_{L\ell}^{*,(m), 1/h} .
\end{equation}
The orthogonal (up to band-limit effects) estimators of this work have $h_\ell = \sqrt{\bar C_\ell} = 1/\sqrt{C^{\rm tot}_\ell}$. For point-source deprojection, we used $h_\ell = I_\ell \sqrt{(2\ell + 1)/(4\pi)^{1/2}}$, see Eq.~\eqref{h_ell_ps}. The response of the estimator to point sources with the expected mean profile is then
\begin{equation}
	\mathcal R_L^{\kappa(m),S^2} =\frac {(2 -\delta_{0m})}{2(2L + 1)} \sqrt{2L + 1}\av{\mathcal T_m^L(\Ml) }_{\Ml} \sum_{\ell}w^{(m), h}_{L\ell}Z_{L\ell} ,
\end{equation}
with $Z_{L \ell} = \frac{1}{2\ell + 1 + L} \frac{A((2\ell -L)/2)}{A((2\ell + L)/2)}$, the $\Ml$-independent part of the Wigner 3j symbol (see Eq.\eqref{eq:3j2}). This response is zero for $m> 0$ up to band-limit effects.
\end{widetext}

\providecommand{\aj}{Astron. J. }\providecommand{\apj}{ApJ
  }\providecommand{\apjl}{ApJ
  }\providecommand{\mnras}{MNRAS}\providecommand{\prl}{PRL}\providecommand{\prd}{PRD}\providecommand{\jcap}{JCAP}\providecommand{\aap}{A\&A}

\end{document}